\documentclass[prx,twocolumn,superscriptaddress,floatfix]{revtex4-2}
\usepackage{bm,graphicx,url,epsf,color,amsmath,amssymb}
\usepackage{MnSymbol}
\usepackage{soul} 

\graphicspath{{Figures/}} 

\begin{document}
\title{Cross-Correlation Investigation of Anyon Statistics in the $\nu=1/3$ and $2/5$ Fractional Quantum Hall States}

\author{P.~Glidic}
\thanks{These authors contributed equally to this work}
\affiliation{Universit\'e Paris-Saclay, CNRS, Centre de Nanosciences et de Nanotechnologies, 91120, Palaiseau, France}
\author{O.~Maillet}
\thanks{These authors contributed equally to this work}
\affiliation{Universit\'e Paris-Saclay, CNRS, Centre de Nanosciences et de Nanotechnologies, 91120, Palaiseau, France}
\author{A.~Aassime}
\affiliation{Universit\'e Paris-Saclay, CNRS, Centre de Nanosciences et de Nanotechnologies, 91120, Palaiseau, France}
\author{C.~Piquard}
\affiliation{Universit\'e Paris-Saclay, CNRS, Centre de Nanosciences et de Nanotechnologies, 91120, Palaiseau, France}
\author{A.~Cavanna}
\affiliation{Universit\'e Paris-Saclay, CNRS, Centre de Nanosciences et de Nanotechnologies, 91120, Palaiseau, France}
\author{U.~Gennser}
\affiliation{Universit\'e Paris-Saclay, CNRS, Centre de Nanosciences et de Nanotechnologies, 91120, Palaiseau, France}
\author{Y.~Jin}
\affiliation{Universit\'e Paris-Saclay, CNRS, Centre de Nanosciences et de Nanotechnologies, 91120, Palaiseau, France}
\author{A.~Anthore}
\email[e-mail: ]{anne.anthore@c2n.upsaclay.fr}
\affiliation{Universit\'e Paris-Saclay, CNRS, Centre de Nanosciences et de Nanotechnologies, 91120, Palaiseau, France}
\affiliation{Universit\'e Paris Cit\'e, CNRS, Centre de Nanosciences et de Nanotechnologies, F-91120 Palaiseau, France}
\author{F.~Pierre}
\email[e-mail: ]{frederic.pierre@cnrs.fr}
\affiliation{Universit\'e Paris-Saclay, CNRS, Centre de Nanosciences et de Nanotechnologies, 91120, Palaiseau, France}

\begin{abstract}
Recent pioneering works have set the stage for exploring anyon braiding statistics from negative current cross-correlations along two intersecting quasiparticle beams.
In such a dual-source - analyzer quantum point contact setup, also referred to as `collider’, the anyon exchange phase of fractional quantum Hall quasiparticles is predicted to be imprinted into the cross-correlations characterized by an effective Fano-factor $P$.
In the case of symmetric incoming quasiparticle beams, conventional fermions result in a vanishing $P$.  
In marked contrast, we observe signatures of anyon statistics in the negative $P$ found both for the $e/3$ Laughlin quasiparticles at filling factor $\nu=1/3$ ($P\approx-2$, corroborating previous findings), and for the $e/5$ quasiparticles in the hierarchical state $\nu=2/5$ ($P\approx-1$).  
Nevertheless, we argue that the quantitative connection between a numerical value of $P\neq0$ and a specific fractional exchange phase is hampered by the influence of the analyzer conductance dependence on the voltages used to generate the quasiparticles. 
Finally, we address the important challenge how to distinguish at $\nu=1/3$ between negative cross-correlations induced by a fractional braid phase, and those resulting from a different Andreev-like mechanism.
Although with symmetric sources $P$ does not exhibit signatures of a crossover when the analyzer is progressively detuned to favor Andreev processes, we demonstrate that changing the balance between sources provides a means to discriminate between the two mechanisms.
\end{abstract}

\maketitle

\section{Introduction}
A variety of exotic quasiparticles are predicted to emerge in low dimensional systems, beyond classification into bosons and fermions 
\cite{Feldman_reviewStatChargeFQH_2021,Pham2000,morel2021fractionalization,lee2020fractint,FuKaneMZM2008,LopesAnyonsNCK2020, moore_nonabel_1991}.
In the archetypal regime of the fractional quantum Hall effect (FQHE), the presence of quasiparticles carrying a fraction of the elementary electron charge $e$ is by now firmly established \cite{goldman_fracQantidot_1995,martin_fracQset_2004,kane1994noneq,Fendley_fraccharge_1995,Saminadayar_fracmeas_1997,de-Picciotto_es3_1997,Reznikov_es5_1999,Dolev_es4_2008,crepieux_pasn_2004,safi_timedeptrspt_2014,Kapfer_FQHE_2019,chamon_noiseLL_1995,roussel_noneqFDT_2016,Bisognin_muwave_2019}.
These quasiparticles are furthermore predicted to exhibit unconventional behaviors upon inter-exchange, different from bosons and fermions, and were accordingly coined any(-)ons \cite{Wilczek1982Anyons}.
Such a possibility results from the topological modification introduced by a double exchange (a braiding) under reduced dimensionalities \cite{Leinaas1977exchange}.
Exchanging two fractional quasiparticles can either add a factor $\exp(i\theta)$ with an exchange phase $\theta$ smaller than the fermionic $\pi$ (Abelian anyons), or result in a drastic change of the wave function not possible to reduce to a simple phase factor (non-Abelian anyons).
Notably, the Laughlin FQHE series at electron filling factor per flux quantum $\nu=1/(2p+1)$ ($p\in\mathbb{N}$) is predicted to host fractional quasiparticles of charge $\nu e$ and exchange phase $\theta=\nu\pi$ as elementary excitations \cite{laughlin1983FQHE,Halperin1984FQHEstat,arovas1984FQHEstat}.
Even more exotic non-Abelian anyons of charge $e/4$ are expected at $\nu=5/2$ \cite{moore_nonabel_1991,Lee_5s2_2007,Levin_Pfaffian_2007} (see \cite{Banerjee2018halfGQ,Dutta_nonAbel_2022} for heat conductance measurements supporting the non-Abelian character).
Providing experimental evidence of a fractional exchange phase proved more challenging than the fractional charge.
It is only recently that first convincing signatures were detected at $\nu=1/3$, from $2\pi/3$ phase jumps in an electronic interferometer \cite{Nakamura_anyon_2020} and through negative cross-correlations in a source-analyzer setup \cite{bartolomei_anyons_2020}.

The two methods are complementary and, specifically, the second \cite{rosenow_collider_2016} promises to be remarkably adaptable to different platforms, including fractional charges propagating along integer quantum Hall channels \cite{morel2021fractionalization,lee2022nonabelian,idrisov2022poscrossnu2}. 
The present work builds upon this source-analyzer approach, by exploring the discerning character of cross-correlation signatures and by expanding the investigation to a different type of anyon. 

We first reexamine the $\nu=1/3$ Laughlin fractional quantum Hall state.
The observations of \cite{bartolomei_anyons_2020} are corroborated over an extended range of analyzer tunings as well as to lower temperatures.
Remarkably, the qualitative signatures of anyon statistics are found to be robust to the setting of the analyzer.
This insensitivity even blurs the frontier with a distinct Andreev-like mechanism \cite{Kane2003Dilu,Glidic2022Andreev1s3} that does not involve an unconventional braid phase.
Nevertheless, we show that it is possible to distinguish anyon braiding by changing the symmetry between sources.
In addition, the remarkable data-theory quantitative agreement previously observed is reproduced here.
However, we show that it relies on a specific normalization choice of the cross-correlation signal.
In essence, extracting direct quantitative information regarding the value of the exchange phase $\theta$, beyond its fractional character, is impeded by the accompanying influence of the analyzer conductance.
Then we investigate the hierarchical (Jain) $\nu=2/5$ state, where $e/5$ quasiparticles are predicted to have a different fractional exchange phase of $3\pi/5$.
The $\nu=2/5$ observation of negative cross-correlations with symmetric sources provides a qualitative signature for the anyon character of these quasiparticles.

\section{Probing anyon statistics with cross-correlations}

\begin{figure}[htb]
	\centering
	\includegraphics[width=8.6cm]{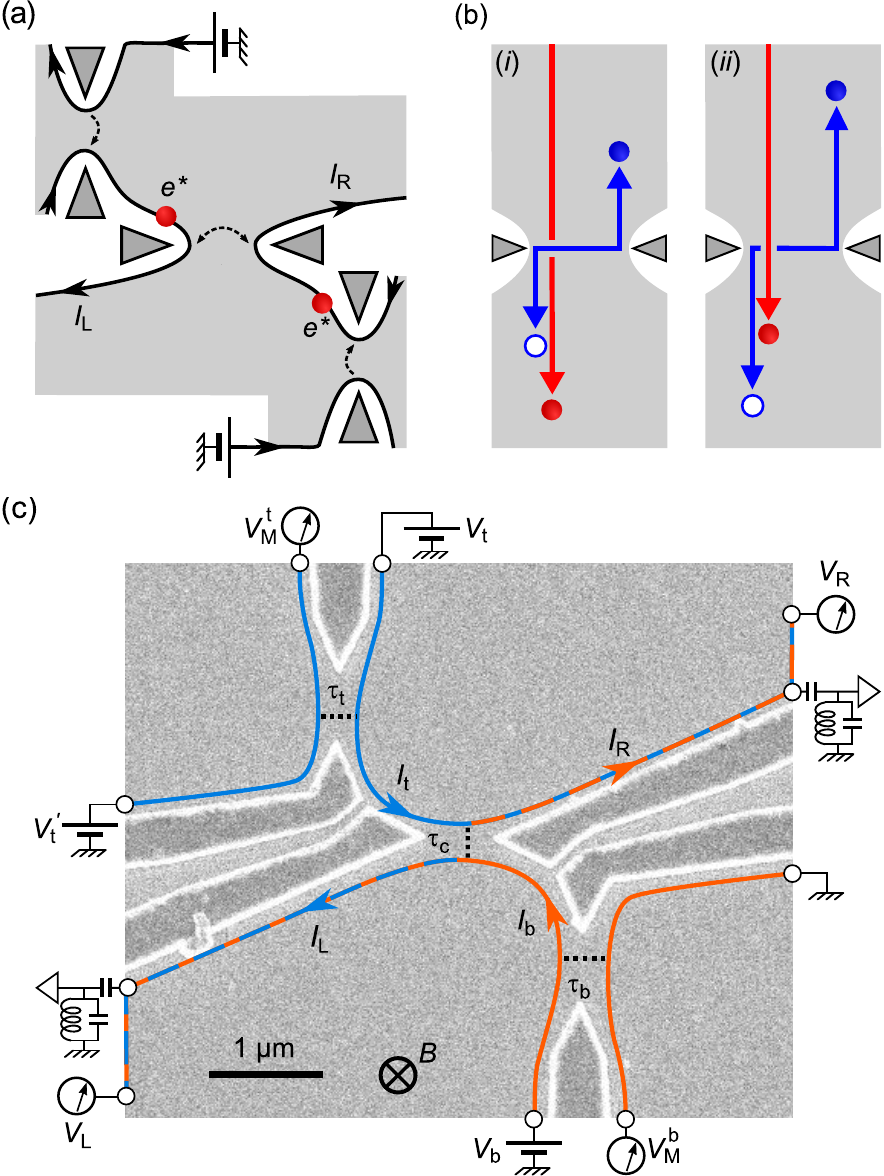}
	\caption{
	(a) Sources-analyzer setup. 
	Quantum point contacts (pairs of facing triangles) at the top-left (QPC$_\mathrm{t}$) and bottom-right (QPC$_\mathrm{b}$) in the weak back-scattering (WBS) regime constitute sources of quasiparticles of fractional charge $e^*$.
	The emitted quasiparticles propagate toward the central `analyzer' QPC$_\mathrm{c}$ along quantum Hall edge channels depicted by lines with arrows (inactive channels not shown).
	Cross-correlations $\langle \delta I_\mathrm{L}\delta I_\mathrm{R}\rangle$ inform on the statistics.
	(b) Braid-induced mechanism. 
	Analyzer tunnelings (double arrow in (a)) result from interferences between the generation of a quasiparticle-quasihole pair across QPC$_\mathrm{c}$ (blue  double-arrow) after \textit{(i)} or before \textit{(ii)} the passing of incident quasiparticles (one represented with red arrow).
	The process cancels for a trivial braid phase $2\pi$.
	(c) Sample ebeam micrograph.
	Metallic gates on the surface of a Ga(Al)As heterojunction appear darker with bright edges.
	QPC$_\mathrm{t,b}$ are tuned to matching transmission ratios  $\tau_\mathrm{t}\approx\tau_\mathrm{b}$ of the active channel.
	The sources imbalance $I_-\equiv I_\mathrm{t}-I_\mathrm{b}$ is controlled with $V_\mathrm{t}-V_\mathrm{b}$.  
	We set $V_\mathrm{t}'=0$ except for the separate shot noise characterization of the central analyzer QPC$_\mathrm{c}$, which is performed with a direct voltage bias ($V_\mathrm{t}'=V_\mathrm{t}$ and $V_\mathrm{b}=0$).
	}
	\label{FigSample}
\end{figure}

The setup probing unconventional anyon statistics is schematically illustrated in Fig.~\ref{FigSample}(a).
It is composed of two random sources of quasiparticles impinging on both sides of a central `analyzer' constriction.
Signatures of unconventional exchange statistics are encoded into the cross-correlations between current fluctuations along the two outgoing paths $\langle \delta I_\mathrm{L}\delta I_\mathrm{R}\rangle$.
In the limit of dilute sources of anyon quasiparticles and of a nearly ballistic short central constriction, theory predicts negative cross-correlations that depend on the balance between the two sources and persist at symmetry \cite{rosenow_collider_2016,morel2021fractionalization,lee2022nonabelian}.
In this section, we first discuss the theoretical origin of the connection between cross-correlations and exotic anyon exchange phase $\theta$.
Then, the discriminating character of this signal, to attest of a non-trivial fractional phase, is assessed by comparing with expectations in different configurations.

How do cross-correlations connect with anyon statistics?
Initially, an intuitive interpretation of the predicted cross-correlations was proposed in terms of a partial bunching of colliding quasiparticles \cite{rosenow_collider_2016}.
However a collision involves two almost simultaneously incoming quasiparticles, and it was recently pointed out that this contribution becomes negligibly small for sources in the considered limit of dilute, randomly emitted quasiparticles \cite{morel2021fractionalization,lee2022nonabelian} (a rapidly diminishing signal, as the square of the dilution ratio, is also expected from a classical model \cite{rosenow_collider_2016}). 
The \textit{same} theoretical prediction was instead attributed to a different interference mechanism, between two different processes labeled (\textit{i}) and (\textit{ii}) in Fig.~\ref{FigSample}(b).
These correspond to the thermal excitation of a quasiparticle-quasihole pair across the analyzer constriction before, or after, the transmission of quasiparticles emitted from the sources \cite{morel2021fractionalization,lee2022nonabelian,jonckheere2022anyonic}.
This is schematically illustrated in Fig.~\ref{FigSample}(b) in the presence of a single incident quasiparticle.
Importantly, such an interference can be mapped onto a braiding between incident and thermally excited anyons \cite{Han2016BubbleBraiding}, and it cancels for a trivial braid phase $2\theta=0 \pmod{2\pi}$.
The pairs generated across the analyzer constriction through this braiding mechanism directly result in a current cross-correlation signal, whose mere existence for symmetric incoming beams constitutes a first marker of unconventional anyon statistics.
Moreover, incident quasiparticles from opposite sources are associated with a braiding along inverse winding directions, and therefore contribute with opposite signs to the relevant total braid phase \cite{morel2021fractionalization,lee2022nonabelian}.
For example, two quasiparticles incident from opposite sides within a time window shorter than $h/k_\mathrm{B}T$ (with $T$ the temperature) are associated with a null total braid phase, leading to a breakdown of this transport mechanism across the analyzer (see \cite{jonckheere2022anyonic} for the detailed dependence in the time delay).
Consequently, the cross-correlations resulting from a non-trivial braiding depend on the balance between the beams of incoming, randomly emitted quasiparticles, which constitutes a second complementary marker.
As recapitulated in Table~\ref{tab:CrossSum}, these two markers combined together provide a strong qualitative signature of an underlying non-trivial anyon statistics.

\begin{table}
\begin{ruledtabular}
\begin{tabular}{ccc|cc}
      \multicolumn{3}{c|}{System}&\multicolumn{2}{c}{Cross-corr.}\\
    
     Platform (mechanism) & $e_\mathrm{t,b}$ & $e_\mathrm{c}$ & Sym.& Asym.  \\
     \colrule
     Laughlin FQHE (braiding)& $\nu e$ & $\nu e$ & $-$ & $--$ \\
     Laughlin FQHE (Andreev)& $\nu e$ & $e$ & $-$& $-$\\
     Free fermions & $e$& $e$ & $0$& $(-)$\\
     Interacting IQHE channels & $e$ & $e$ & $+$& $(-)$\\
\end{tabular}
\end{ruledtabular}
\caption{\label{tab:CrossSum}{
Cross-correlations with dilute beams of incident quasiparticles. 
Both the cross-correlation sign and evolution between symmetric sources (Sym.)\ and a single source (Asym.)\ are compulsory to distinguish between different transport mechanisms involving tunneling quasiparticles of charge $e_\mathrm{t,b,c}$.
Parentheses indicate a signal that emerge for non-dilute incident beams and a $--$ signifies a larger amplitude.
See \cite{idrisov2022poscrossnu2} for the predictions of positive cross-correlations with two interacting channels of the integer quantum Hall effect (IQHE), and \cite{Kane2003Dilu,Glidic2022Andreev1s3} for the prediction and observation of an Andreev mechanism giving rise to symmetry-independent negative cross-correlations when the analyzer is set to favor the tunneling of quasielectrons.
}
}  

\end{table}

\section{Experimental implementation}

The device shown in Fig.~\ref{FigSample}(c) is realized on a Ga(Al)As two-dimensional electron gas of density $1.2\times10^{11}$\,cm$^{-2}$ located 140\,nm below the surface.
It is cooled at a temperature $T\simeq35\,$mK (if not stated otherwise) and immersed in a strong perpendicular magnetic field $B$ corresponding to the middle of the quantum Hall effect plateau at filling factors $\nu=1/3,$ $2/5$ and 2 (see Appendix~C for $\nu=2$).
In the quantum Hall regime, the current flows along chiral edge channels depicted as lines with arrows indicating the propagation direction.
At $\nu=2/5$ and 2 there are two co-propagating quantum Hall channels with the same chirality, although, for clarity, only the active one in which non-equilibrium quasiparticles are injected is displayed in Fig.~\ref{FigSample}.

The sources and analyzer constrictions are realized by voltage biased quantum point contacts (QPCs) tuned by field effect using metal split gates (darker with bright edges).
The source QPCs located in the top-left and bottom-right of Fig.~\ref{FigSample}(c) are referred to as QPC$_\mathrm{t}$ and QPC$_\mathrm{b}$, respectively.
The central analyzer QPC is referred to as QPC$_\mathrm{c}$.
The sources are connected to the downstream QPC$_\mathrm{c}$ by an edge path of approximately $1.5\,\mu$m.

In the following, we first discuss the characterization of the current fraction going through the source and analyzer.
Then, we detail the determination of the fractional charges of the tunneling quasiparticles, and whether this characterization can be performed simultaneously with the measurement of the main cross-correlation signal or separately. 

\subsection{QPC transmission}
QPC$_\mathrm{t,b,c}$ are first characterized through the fractions $\tau_\mathrm{t,b,c}$ of (differential) current in the active channel transmitted across the constriction:
\begin{align}
\tau_\mathrm{t(b)} &\equiv \frac{\nu}{\nu_\mathrm{eff}}\left(\frac{ \partial V_\mathrm{M}^\mathrm{t(b)}}{\partial V_\mathrm{t(b)}}-1\right)+1,
\label{EqTauS} \\
\tau_\mathrm{c} &\equiv \frac{\nu}{\nu_\mathrm{eff}}\left( \frac{\partial V_\mathrm{R}/\partial V_\mathrm{b}}{2(1-\tau_\mathrm{b})}+\frac{\partial V_\mathrm{L}/\partial V_\mathrm{t}}{2(1-\tau_\mathrm{t})}\right),
\label{EqTauC}
\end{align}
with the partial derivatives given by lock-in measurements, and where $\nu_\mathrm{eff}$ is the effective filling factor associated with the conductance $\nu_\mathrm{eff}e^2/h$ of the active channel ($\nu_\mathrm{eff}=\nu$ if there is a single channel, $\nu_\mathrm{eff}=1/15$ for the inner channel at $\nu=2/5$, and $\nu_\mathrm{eff}=1$ for each channel at $\nu=2$).
Note that we follow the standard convention for the definition of the transmission direction across the QPCs' split gates, as indicated by dashed lines in Fig.~\ref{FigSample}(c).
The so-called strong back-scattering (SBS) and weak back-scattering (WBS) regimes correspond to $\tau\ll1$ and $1-\tau\ll1$, respectively.
As discussed below and illustrated in Fig.~\ref{FigSample}(a), the sources and analyzer are normally set in the WBS regime to emit and probe the statistics of fractional quasiparticles.
The top-right inset in Fig.~\ref{Fig1s3repres}(c) displays such $\tau_\mathrm{t,b,c}$ measurements.

\subsection{Quasiparticle sources}
Applying a voltage bias $V_\mathrm{t(b)}$ excites the quantum Hall edge channel at the level of QPC$_\mathrm{t(b)}$ (except for $\tau_\mathrm{t(b)}\in\{0,1\}$), hence generating a quasiparticle carrying current $I_\mathrm{t(b)}$ propagating toward the analyzer.
The nature of these quasiparticles depends on the tuning of the QPCs.
For Laughlin fractions $\nu$, their charge is predicted to be $e$ at $\tau\ll1$ and $\nu e$ at $1-\tau\ll1$ \cite{kane1994noneq,Fendley_fraccharge_1995}.
We characterize the charge $e_\mathrm{t(b)}$ of the quasiparticles emitted at QPC$_\mathrm{t(b)}$ by confronting the fluctuations of $I_\mathrm{t(b)}$ with the standard, phenomenological expression for the excess shot noise \cite{blanter_shotnoise_2000,Feldman_SNfrac_2017}:
\begin{equation}
   \langle\delta I^2\rangle_\mathrm{exc}=2e^*\tau^\mathrm{dc}(1-\tau^\mathrm{dc})\frac{\nu_\mathrm{eff}e^2}{h}V\left[\coth{\frac{e^*V}{2k_{\mathrm{B}}T}}-\frac{2k_{\mathrm{B}}T}{e^*V}\right],
    \label{EqSN}
\end{equation}
where $\delta I\equiv I-\langle I\rangle$, $\langle\delta I^2\rangle_\mathrm{exc}\equiv\langle\delta I^2\rangle(V)-\langle\delta I^2\rangle(0)$ , $I=I_\mathrm{t(b)}$, $e^*=e_\mathrm{t(b)}$, $V=V_\mathrm{t(b)}$ and $\tau^\mathrm{dc}$ an alternative definition of $\tau_\mathrm{t(b)}$ with the derivative in Eq.~\ref{EqTauS} replaced by the ratio of the dc voltages.
Note that the charge $e^*$ is extracted focusing on $e^*V\gg k_\mathrm{B}T$, while the $\coth{}$ transient is only a rough approximation to the predicted low voltage behavior \cite{Snizhko2015scaldimFano,Schiller2022scaldimFano}.
In practice, we measure the auto- and cross-correlations of $\delta I_\mathrm{L,R}$ and not directly the current fluctuations emitted by the sources.
The main approach here used to determine the shot noise from the sources is to consider the measured noise sum defined as:
\begin{equation}
    S_\Sigma\equiv \langle \delta I_\mathrm{L}^2\rangle_\mathrm{exc}+\langle \delta I_\mathrm{R}^2\rangle_\mathrm{exc}+2\langle \delta I_\mathrm{L}\delta I_\mathrm{R}\rangle.
    \label{EqSsigma}
\end{equation}
Current conservation ($I_\mathrm{t}+I_\mathrm{b}=I_\mathrm{L}+I_\mathrm{R}$) together with the absence of current correlations between sources expected from chirality ($\langle \delta I_\mathrm{t}\delta I_\mathrm{b}\rangle=0$) imply:
\begin{equation}
    \langle\delta I_\mathrm{t}^2\rangle_\mathrm{exc}+\langle\delta I_\mathrm{b}^2\rangle_\mathrm{exc}=S_\Sigma.
    \label{EqSsigmavsSourceSN}
\end{equation}
This approach is systematically used simultaneously with the measurement of the anyon statistics cross-correlation signal. 
With two active sources in this case (both $V_\mathrm{t,b}\neq0$), $S_\Sigma$ informs us on the weighted average of $e_\mathrm{t}$ and $e_\mathrm{b}$ (see e.g.\ Fig.~\ref{Fig1s3repres}(a)).
Such an approach is also applied with a single active source, sweeping $V_\mathrm{t(b)}$ at fixed $V_\mathrm{b(t)}=0$.
For perfectly independent sources, the increase of $S_\Sigma$ then corresponds to the excess shot noise across QPC$_\mathrm{t(b)}$, providing us separately with the quasiparticles' charge $e_\mathrm{t(b)}$.
(As discussed later, some imperfections may however develop.)
Note that, when possible, we check the consistency of the extracted charges $e_\mathrm{t,b}$ with the values obtained by setting the analyzer to a full transmission or a full reflection ($\tau_\mathrm{c}\in\{0,1\}$), where there is a straightforward one-to-one correspondence between $I_\mathrm{t,b}$ and $I_\mathrm{L,R}$.

\subsection{Analyzer tunneling charge}
The individual shot noise characterization of QPC$_\mathrm{c}$ requires the application of a direct voltage bias, as opposed to incident currents composed of non-equilibrium quasiparticles.
Hence, it must be performed in a dedicated, separate measurement.
In practice, we set $V_\mathrm{t}'=V_\mathrm{t}$ at $V_\mathrm{b}=0$ (see Fig.~\ref{FigSample}(c), elsewhere $V_\mathrm{t}'=0$) without changing the gate voltage tuning of any of the QPCs, and we measure the resulting cross-correlations $\langle\delta I_\mathrm{L}\delta I_\mathrm{R}\rangle$ (see Fig.~\ref{fig:AutoVsCross} in Appendix~E for the less robust auto-correlation signal).
Fitting the noise slopes with the negative of the prediction of Eq.~\ref{EqSN} provides us with the characteristic charge $e_\mathrm{c}$ of the quasiparticles transmitted across QPC$_\mathrm{c}$ (see Fig.~\ref{Fig1s3repres}(b) and Fig.~\ref{Fig2s5repres}(b)).

\subsection{Experimental procedure}
With these tools, the device is set to have two sources of transmission probabilities that remain symmetric $\tau_\mathrm{t}\approx\tau_\mathrm{b}$ and with the same fractional quasiparticle charges $e_\mathrm{t}\simeq e_\mathrm{b}\simeq e^*$, over the explored range of bias voltages of typically $V_\mathrm{t,b}\lesssim100\,\mu$V.
At $\nu=1/3$, $2/5$ and 2, we focus on $e^*/e\simeq1/3$, $1/5$ and 1, respectively.
The symmetry between the two quasiparticle beams impinging on the analyzer is then controlled through $V_\mathrm{t}$ and $V_\mathrm{b}$, and characterized by the ratio $|I_-/I_+|$ with $I_\pm\equiv I_\mathrm{t}\pm I_\mathrm{b}$.
The analyzer QPC$_\mathrm{c}$ is normally set to the same tunneling charge $e_\mathrm{c}\simeq e^*$ to investigate the fractional exchange phase of $e^*$ quasiparticles, although a broader range of $e_\mathrm{c}$ is also explored at $\nu=1/3$ by tuning the analyzer QPC$_\mathrm{c}$ away from the WBS regime.

\section{Theoretical predictions\label{SectionThyPred}}

We recapitulate the cross-correlation predictions for free electrons \cite{blanter_shotnoise_2000} and anyons of the Laughlin series \cite{rosenow_collider_2016,Lee_NegExcSN_2019,lee2022nonabelian}. 
Other related theoretical developments include the recent extensions to co-propagating integer quantum Hall channels in interactions \cite{idrisov2022poscrossnu2}, to fractional charge injected in integer quantum Hall channels \cite{morel2021fractionalization}, to non-Abelian anyons \cite{lee2022nonabelian}, to high frequencies \cite{safi_fdtnoneq_2020} and to Laughlin quasiparticles with a controlled time delay \cite{jonckheere2022anyonic}.

\subsection{Effective Fano factor}
The statistics is specifically investigated through the effective Fano factor $P$ defined as:
\begin{equation}
    \label{EqP}
    P \equiv \frac{\langle \delta I_\mathrm{L}\delta I_\mathrm{R}\rangle}{S_\Sigma \tau_\mathrm{c}(1-\tau_\mathrm{c})},
\end{equation}
with a denominator chosen to minimize the direct, voltage-dependent contribution of the shot noise from the sources, thus focusing on the signal of interest generated at the analyzer.
This expression generalizes the definition of $P$ introduced in \cite{rosenow_collider_2016} beyond the asymptotic limits $1-\tau_\mathrm{t,b,c}\ll1$ (where at large bias $S_\Sigma\approx 2e^*I_+$), in the same spirit as in \cite{bartolomei_anyons_2020}.
Note that $\tau_\mathrm{c}$ in the denominator remains the simultaneously measured differential transmission probability given by Eq.~\ref{EqTauC} including in the presence of asymmetric incident quasiparticle beams. 
This is in contrast to \cite{rosenow_collider_2016} with quantitative consequences for asymmetric sources as further discussed in \ref{SectionAnyons}.

\subsection{Fermions}
In the Landauer-B{\"u}ttiker framework for non-interacting electrons, the cross-correlations can be written as \cite{blanter_shotnoise_2000}:
\begin{equation}
    \label{EqCrossfree}
    \langle \delta I_\mathrm{L}\delta I_\mathrm{R}\rangle =  -2\tau_\mathrm{c}(1-\tau_\mathrm{c})(e^2/h) \int d\epsilon \left[f_\mathrm{t}(\epsilon)-f_\mathrm{b}(\epsilon) \right]^2,
\end{equation}
where $f_\mathrm{t,b}$ are the energy distribution functions of electrons incoming on QPC$_\mathrm{c}$ from the top (t) and bottom (b) paths.
The cross-correlations and, consequently, $P$ are thus expected to robustly vanish in the symmetric limit, whenever $f_\mathrm{t}\simeq f_\mathrm{b}$ (positive cross-correlations are expected within the bosonic density wave picture emerging for interacting, adjacent integer quantum Hall channels \cite{idrisov2022poscrossnu2}).
Furthermore, in the dilute incident beam limit where $|f_\mathrm{t}-f_\mathrm{b}|\ll1$, $P$ remains asymptotically null even in the presence of an asymmetry.
In this limit and for symmetric configurations, the contrast is thus particularly striking with the cross-correlations predicted for anyons.

\subsection{Anyons\label{SectionAnyons}} 
Theoretical solutions for the sources-analyzer setup were obtained for Laughlin fractions $\nu=1/(2p+1)$ at low temperatures ($e^*V_\mathrm{t,b}\gg k_\mathrm{B}T$), in the WBS regime of the source QPC$_\mathrm{t,b}$ ($1-\tau_\mathrm{t,b}\ll1$), and in both the WBS ($1-\tau_\mathrm{c}\ll1$) \cite{rosenow_collider_2016} and SBS ($\tau_\mathrm{c}\ll1$) \cite{Kane2003Dilu} regimes for the analyzer QPC$_\mathrm{c}$.\\

\noindent\textit{Braiding} --
We consider configurations with all QPCs in the WBS regime ($1-\tau_\mathrm{t,b,c}\ll1$), where the occurrence of a non-trivial fractional exchange phase $\theta$ between anyons is predicted to play a crucial role \cite{rosenow_collider_2016,Lee_NegExcSN_2019,lee2022nonabelian}.
The prediction for the effective Fano factor $P$ defined in Eq.~\ref{EqP} with $\tau_\mathrm{c}$ given by Eq.~\ref{EqTauC} reads \cite{rosenow_collider_2016}:
\begin{align}
    P_\mathrm{thy}^\mathrm{WBS}(I_-/I_+)=
    &-4\Delta/(1-4\Delta)\label{EqPBraid}\\
    &+|I_-/I_+|\left\{\left(\tan{2\pi\Delta}+\frac{(1-4\Delta)^{-1}}{\tan{2\pi\Delta}}\right)\right.\nonumber\\
    &\left. \times\tan{\left[(4\Delta-2)\arctan{\frac{|I_-/I_+|}{\tan{2\pi\Delta}}}\right]}\right\},\nonumber
\end{align}
with $I_\pm\equiv I_\mathrm{t}\pm I_\mathrm{b}$ and $\Delta$ the quasiparticles' scaling dimension, which is related to the exchange phase through $\theta=2\pi\Delta$ and, for Laughlin fractions, given by $\Delta=\nu/2$ (see e.g.\ \cite{Schiller2022scaldimFano}).
Note that the above formulation ignores possible non-universal complications, such as edge reconstruction (see \cite{rosenow_collider_2016} for a discussion of such artifacts, and \cite{lee2022nonabelian} for an alternative formulation of $P_\mathrm{thy}^\mathrm{WBS}$ at $I_-=0$ separating the different contributions of braiding phase, topological spin and tunneling exponent).
In the symmetric case ($I_-=0$), Eq.~\ref{EqPBraid} simplifies into $P_\mathrm{thy}^\mathrm{WBS}(0)=-4\Delta/(1-4\Delta)$, which gives $P_\mathrm{thy}^\mathrm{WBS}(0)=-2$ at $\nu=1/3$ and progressively less negative values for lower $\nu$ in the Laughlin series.

As mentioned below Eq.~\ref{EqP}, the dependence in $I_-/I_+$ of $P_\mathrm{thy}^\mathrm{WBS}$ is different from \cite{rosenow_collider_2016}. 
This stems from a different definition for $\tau_\mathrm{c}$ at $I_-\neq0$.
Indeed, multiple definitions are possible for $\tau_\mathrm{c}$, corresponding to different quantitative predictions for this transmission and, consequently, a different $P$  (see Table~\ref{tab:FanoValues}).
In particular, it could be defined as the differential transmission of the current originating from the electrode voltage biased at $V_\mathrm{t,b}$ ($\tau_\mathrm{c}$ given by Eq.~\ref{EqTauC}) or from the other, grounded electrode feeding the source QPCs ($\tau_\mathrm{c}^\mathrm{ter}$ in Table~\ref{tab:FanoValues}).
With sources in the WBS regime, the former $\tau_\mathrm{c}$ corresponds to the transmission across the analyzer of dilute quasiparticles of energy $\sim e^*V$, whereas the latter $\tau_\mathrm{c}^\mathrm{ter}$ is essentially the transmission of a thermal current. 
As the transmission is predicted to strongly depend on energy in the FQHE regime, using these different definitions in Eq.~\ref{EqP} for $P$ clearly results in strongly different theoretical values, as summarized in Table~\ref{tab:FanoValues} (see also Fig.~\ref{Fig1s3PvsAsym}).
Note that in \cite{rosenow_collider_2016}, the alternative definition $\tau_\mathrm{c}^\mathrm{bis}$ relies on the same expression Eq.~\ref{EqTauC} but for $I_-=0$, even for cross-correlations measured at $I_-\neq0$.
As $\tau_\mathrm{c}$ is expected to depend on $I_-$, this impacts the prediction for $P(I_-/I_+\neq0)$.\\ 

\begin{table}
\begin{ruledtabular}
\begin{tabular}{ccc}
     $\tau_\mathrm{c}$ variants & \multicolumn{2}{c}{$P_\mathrm{thy}^\mathrm{WBS}$}\\
     & $(I_-\ll I_+)$ &  $(I_-=I_+)$ \\
     \colrule
    $\tau_\mathrm{c}(I_-,I_+)$ from Eq.~\ref{EqTauC} & $-2$ & $-4.9$ \\
    $\tau_\mathrm{c}^\mathrm{bis}\equiv\tau_\mathrm{c}(I_-=0,I_+)$ \cite{rosenow_collider_2016}& $-2$ & $ -3.1$ \\
     $\tau_\mathrm{c}^\mathrm{ter}\equiv \tau_\mathrm{t}^{-1}\partial V_\mathrm{L}/\partial V_\mathrm{t}'$  \cite{lee2022nonabelian}& $-0.8$ & $-1.3$ \\
\end{tabular}
\end{ruledtabular}
\caption{\label{tab:FanoValues}{
Predicted $P_\mathrm{thy}^\mathrm{WBS}$ at $\nu=1/3$ for alternative definitions of $\tau_\mathrm{c}$ and different sources settings (symmetric when $I_-=0$ or fully asymmetric when $I_-=\pm I_+$, with $I_\pm\equiv I_\mathrm{t}\pm I_\mathrm{b}$).
$\tau_\mathrm{c}$ is the transmission ratio of incident quasiparticles, $\tau_\mathrm{c}^\mathrm{bis}$ the same transmission ratio but at $I_-=0$,
and $\tau_\mathrm{c}^\mathrm{ter}$ the transmission ratio of thermal excitations.
The corresponding values of $P_\mathrm{thy}^\mathrm{WBS}$ are obtained, respectively, from Eq.~\ref{EqPBraid}, and Eqs.~\ref{EqPbis}, \ref{EqPter} in Appendix~F.
}}  
\end{table}

\begin{figure*}[htb]
	\centering
	\includegraphics[width=17.6cm]
	{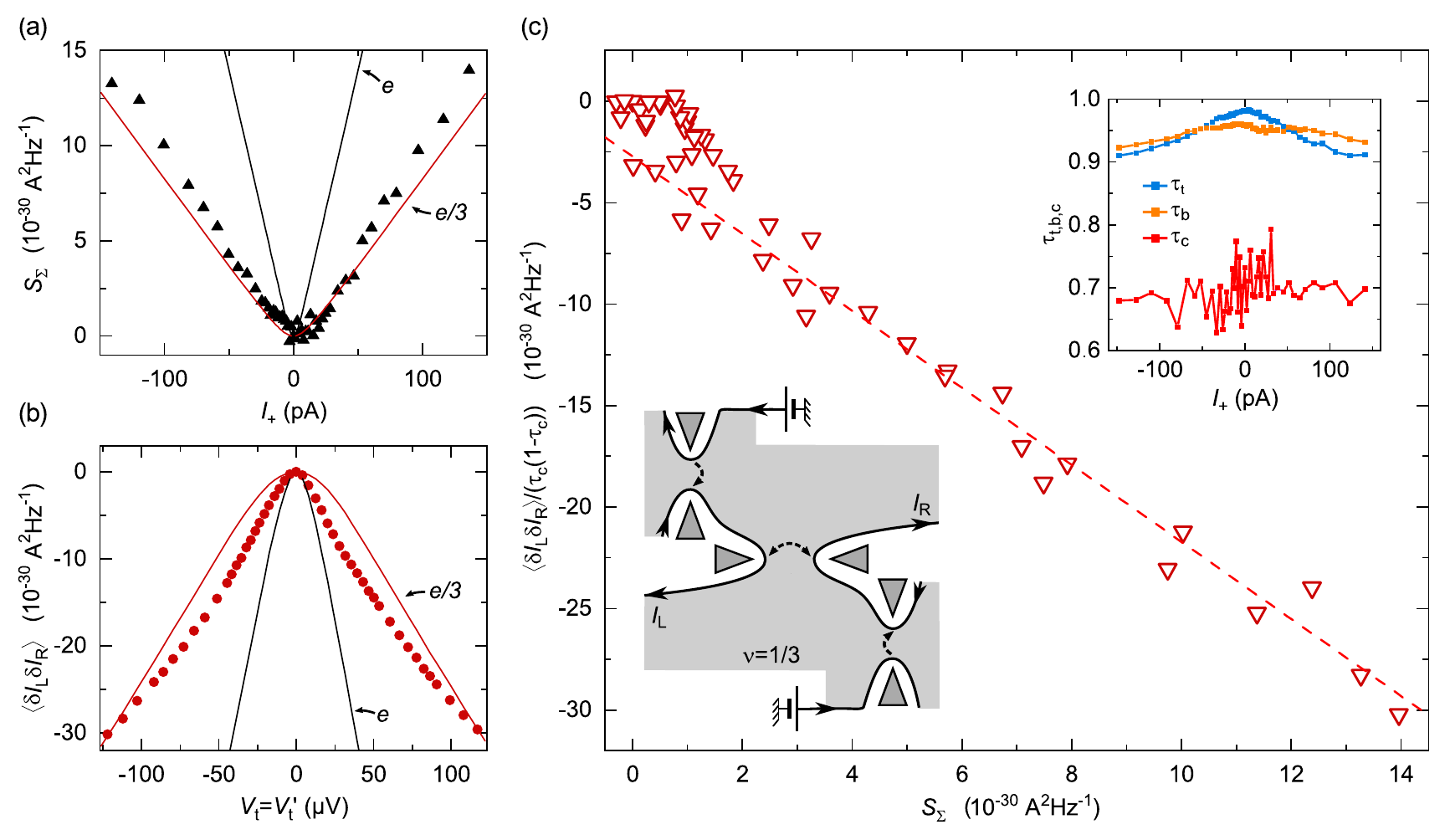}
	\caption{Cross-correlation signature of anyons at $\nu=1/3$ with symmetric sources. 
	All QPCs are in the WBS regime ($\tau_\mathrm{t,b}\approx0.96$, $\tau_\mathrm{c}\simeq0.7$, see schematic illustration and inset in (c)).
	(a),(b) Shot noise characterization of (a) the charge of the quasiparticles emitted from the sources QPC$_\mathrm{t,b}$ ($I_+\equiv I_\mathrm{t}+I_\mathrm{b}$, $S_\Sigma$ given by Eq.~\ref{EqSsigma}), and (b) the tunneling charge across the analyzer QPC$_\mathrm{c}$.
	A source bias $V_\mathrm{t}=V_\mathrm{b}$ ($V_\mathrm{t}'=0$) is applied for the simultaneous measurements in (a) and (c), whereas $V_\mathrm{t}=V_\mathrm{t}'$ ($V_\mathrm{b}=0$) implements a direct voltage bias of QPC$_\mathrm{c}$ in (b).
	The noise data (symbols) match the predictions for $e/3$ (red lines).
	(c) Cross-correlations in the symmetric sources-analyzer configuration.
	The effective Fano factor $P$ is obtained from a linear fit (dashed line) of the slope of the normalized cross-correlation data (symbols) plotted as a function of $S_\Sigma$.
	Here $P\simeq-1.9$. 
	Inset: QPCs transmission.
	}
	\label{Fig1s3repres}
\end{figure*}

\noindent\textit{Andreev reflection} --
We consider here the `Andreev' configuration where QPC$_\mathrm{t,b}$ remain in the WBS regime while QPC$_\mathrm{c}$ is set in the SBS regime ($\tau_\mathrm{c}\ll1$).
In this case, quasielectrons of charge $e$ are tunneling across QPC$_\mathrm{c}$ \cite{kane1994noneq,Fendley_fraccharge_1995,Griffiths2000QPStoWBS,Glidic2022Andreev1s3}.
As the braid phase between such a quasielectron and the impinging fractional quasiparticles is a trivial $2\pi$ for the Laughlin quantum Hall fractions \cite{lee2020fractint,Schiller2022scaldimFano}, the previously discussed transport mechanism driven by unconventional anyon statistics here cancels out.
Instead, a different Andreev-like process takes place, involving independent tunnelings of $e$ accompanied by the simultaneous reflection of a hole of charge $-e(1-\nu)$ \cite{Kane2003Dilu}, as recently observed at $\nu=1/3$ \cite{Glidic2022Andreev1s3}.
As a result, the cross-correlations are simply $-(1-\nu)$ times the shot noise on the tunneling current given by $2 e I_+ \tau_\mathrm{c}$ and, at high bias $\nu eV\gg k_\mathrm{B}T$, $P_\mathrm{thy}^\mathrm{SBS}\simeq -(1-\nu)/\nu$ independently of the ratio $I_-/I_+$ \cite{Kane2003Dilu}.
At $\nu=1/3$, this gives $P_\mathrm{thy}^\mathrm{SBS}\simeq-2$ identical to the exchange induced prediction at symmetry $P_\mathrm{thy}^\mathrm{WBS}(0)=-2$.
Note that this matching is specific to $\nu=1/3$, and does not apply to other Laughlin fractions.
Importantly, a qualitatively distinctive feature of the Andreev-like process is its independence in $I_-/I_+$ \cite{Kane2003Dilu,Glidic2022Andreev1s3}.

\section{Anyon signatures at $\nu=1/3$}

\subsection{Representative anyon signature with symmetric sources}
Figure~\ref{Fig1s3repres} displays some of the QPCs characterization data as well as the cross-correlation anyon signature for a representative WBS device tuning, with symmetric sources ($I_-/I_+\ll1$) at $T\simeq35\,$mK.
The data shown as symbols in panels (a) and (c) are measured simultaneously, whereas the shot noise characterization of QPC$_\mathrm{c}$ shown in (b) is performed separately, slightly before, as it involves a direct voltage bias of the analyzer.

The source QPC$_\mathrm{t,b}$ are set in the WBS limit, at $1-\tau_\mathrm{t,b}<0.1$ for all the data in this section (see illustrative schematic and inset in Fig.~\ref{Fig1s3repres}(c)).
The charge $e_\mathrm{t,b}\approx e/3$ of the emitted quasiparticles is attested by the comparison in Fig.~\ref{Fig1s3repres}(a) with the standard shot noise expression Eq.~\ref{EqSN}. 
The measured noise sum $S_\Sigma$ (Eq.~\ref{EqSsigma}), corresponding to the shot noise from both sources, is found in close agreement with $e^*=e/3$ at $T=35$\,mK (a similar matching is obtained from individual source characterizations performed separately).
Note that we limit the applied bias voltage to $|V_\mathrm{t,b}|\lesssim100\,\mu$V. 

\begin{figure}[htb]
	\centering
	\includegraphics[width=8.6cm]
	{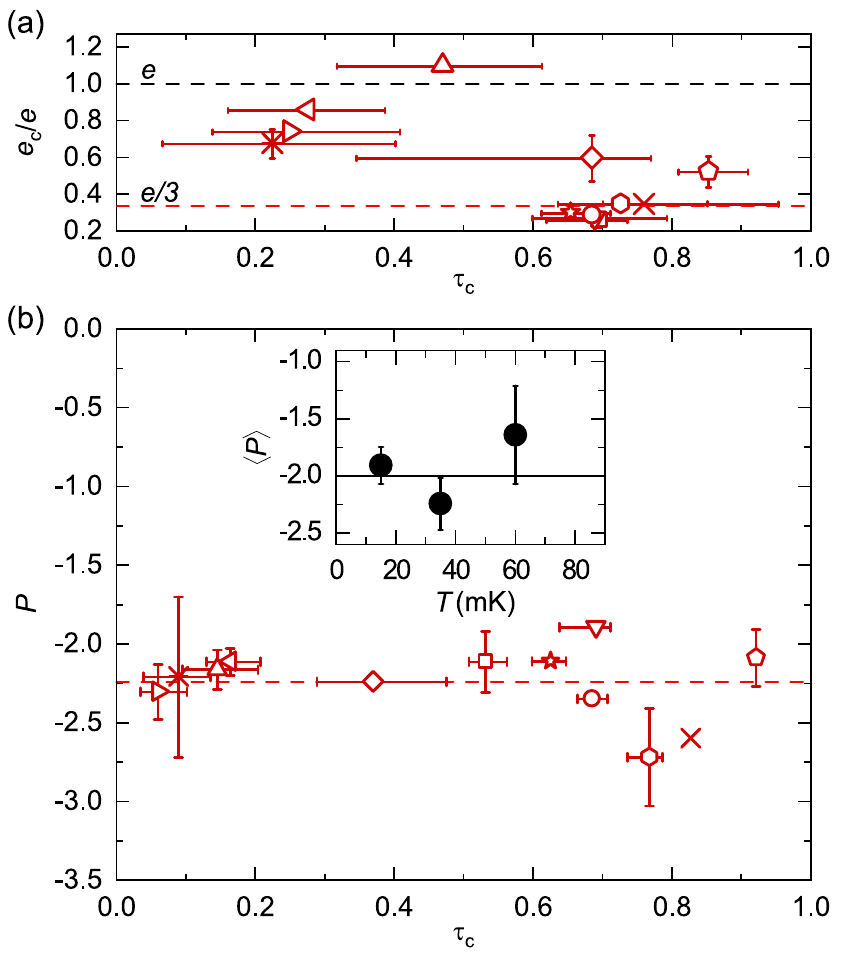}
	\caption{$P(I_-\approx0)$ vs analyzer tuning at $\nu=1/3$.
	Identical symbols represent measurements for the same device tuning ($\tau_\mathrm{c}$ differs in (a) and (b) because of the different biasing).
	Error bars are shown if larger than symbols.
	The horizontal error bars encompass the variation of $\tau_\mathrm{c}$ in the range of biases where $e_\mathrm{c}$ (a) or $P$ (b) are extracted.
	The vertical error bars encompass the difference between measurements at negative and positive voltages.
	(a) Analyzer crossover from $e_\mathrm{c}\approx e/3$ to $e$. 
	(b) Indiscernible crossover in $P(I_-\approx0,\tau_\mathrm{c})$. 
	The horizontal dashed line displays the mean value $\langle P\rangle\simeq-2.2$.
	Inset: mean value $\langle P\rangle$ vs temperature. 
	The vertical error bars show the standard deviation between values of $P$ for individual analyzer settings.
	}
	\label{Fig1s3robustness}
\end{figure}

The analyzer QPC$_\mathrm{c}$ transmission $\tau_\mathrm{c}\approx0.7$ simultaneously measured in the source-analyzer configuration, with impinging dilute quasiparticle beams, is shown in the inset of Fig.~\ref{Fig1s3repres}(c).
The larger experimental noise, particularly marked at low $I_+$, simply reflects the lower amount of current probing $\tau_\mathrm{c}$ at the corresponding low values of $1-\tau_\mathrm{t,b}$. 
The shot noise characterization in Fig.~\ref{Fig1s3repres}(b) is separately performed from the cross-correlations measured in the presence of a direct voltage bias (see Appendix E for auto-correlations and $S_\Sigma$ data).
A good agreement is observed with the negative of Eq.~\ref{EqSN} for $e^*=e/3$ and $T=35$\,mK.
Note that at relatively low voltages ($e^*V\sim k_\mathrm{B}T$), the data exhibit a noticeably larger slope than the phenomenological Eq.~\ref{EqSN}, which is expected from exact predictions for the thermal rounding \cite{Snizhko2015scaldimFano,Schiller2022scaldimFano}. 
In practice, following standard procedures, we extract the tunneling charge $e_\mathrm{c}$ by fitting the cross-correlations at voltages above the thermal rounding ($e_\mathrm{c}V\gtrsim 3 k_\mathrm{B}T$, with here $e_\mathrm{c}\simeq0.30e$, see also Fig.~\ref{Fig1s3robustness}(a)).

The main cross-correlation signal in the presence of symmetric beams of incident quasiparticles, is normalized by $\tau_\mathrm{c}(1-\tau_\mathrm{c})$ and plotted in the main panel in Fig.~\ref{Fig1s3repres}(c) as a function of the noise sum $S_\Sigma$.
In this representation, the experimental value of $P$ is straightforwardly obtained from a linear fit of the data.
The dashed line corresponds to $P\simeq-1.9$.

Although the quantitative agreement with the prediction $P_\mathrm{thy}^\mathrm{WBS}(0)=-2$ is striking, and corroborates the pioneer observation \cite{bartolomei_anyons_2020}, it is nevertheless counterbalanced by the strong dependence of $P_\mathrm{thy}^\mathrm{WBS}$ on the specific definition of $\tau_\mathrm{c}$ (see Table~\ref{tab:FanoValues}).
This is in contrast with the weakly dependent experimental $P$.
Indeed, we observe in practice $\tau_\mathrm{c}\sim\tau_\mathrm{c}^\mathrm{bis}\sim\tau_\mathrm{c}^\mathrm{ter}$ (with discrepancies smaller than 10\%, of the order of our in-situ experimental resolution on $\tau_\mathrm{c}$) thus leaving $P$ mostly unchanged, as opposed to the different predictions.
This situation can be traced back to the bias voltage dependence of $\tau_\mathrm{c}$, that sharply differs from the expected power-law $1-\tau_\mathrm{c}\propto V^{2\nu-2}$ \cite{Kane&Fisher1992PRL} (see inset in Fig.~\ref{Fig1s3repres}(c) for a representative weak dependence of $\tau_\mathrm{c}$, and also Appendix~E for a measurement as a function of a direct voltage bias).
Nevertheless, the qualitative observation of a non-zero, negative $P$ in the WBS regime with symmetric quasiparticle beams remains a significant, robust feature.
This constitutes in itself a key marker of unconventional exchange statistics.

\subsection{Intriguing robustness of $P(I_-\approx0)$ vs analyzer tuning}
Figure~\ref{Fig1s3robustness} synthesizes our measurements of $P$ at $\nu=1/3$ while broadly changing the tuning of QPC$_\mathrm{c}$ from WBS to SBS (with the sources remaining in the WBS regime and symmetric, $I_-\ll I_+$).
As detailed below, whereas the predicted underlying mechanism changes from anyon braiding to Andreev, no signature of this crossover is discernible in $P(I_-\approx0,\tau_\mathrm{c})$. 
Although there is no contradiction with theory, this calls for additional ways to directly distinguish the two mechanisms. 

The crossover from $e_\mathrm{c}\approx e/3$ to $e$ as $\tau_\mathrm{c}$ is reduced is established in Fig.~\ref{Fig1s3robustness}(a) \cite{kane1994noneq,Fendley_fraccharge_1995,Griffiths2000QPStoWBS}.
Accordingly, a different Andreev transport mechanism is expected to dominate at $\tau_\mathrm{c}\lesssim0.5$, as predicted \cite{Kane2003Dilu} and recently observed on the same sample \cite{Glidic2022Andreev1s3}.
Although an identical value $P=-2$ is asymptotically predicted for both WBS and SBS tunings of the analyzer, signatures of the crossover between different underlying mechanism could have emerged at intermediate $\tau_\mathrm{c}$.
This is not the case.
Instead a remarkable robustness of $P$ versus $\tau_\mathrm{c}$ is observed, as shown in the main panel of Fig.~\ref{Fig1s3robustness}(b) for $T\simeq35\,$mK.
This observation is confirmed at $T\simeq15$\,mK and 60\,mK as can be inferred from the mean and standard deviation of $P$ displayed for different $T$ in the inset.
With no signature of a change of an underlying mechanism materializing along the crossover from WBS to SBS, it is highly desirable to have a direct signature that differentiates between braiding and Andreev processes.

\begin{figure}[t!]
	\centering
	\includegraphics[width=8.6cm]
	{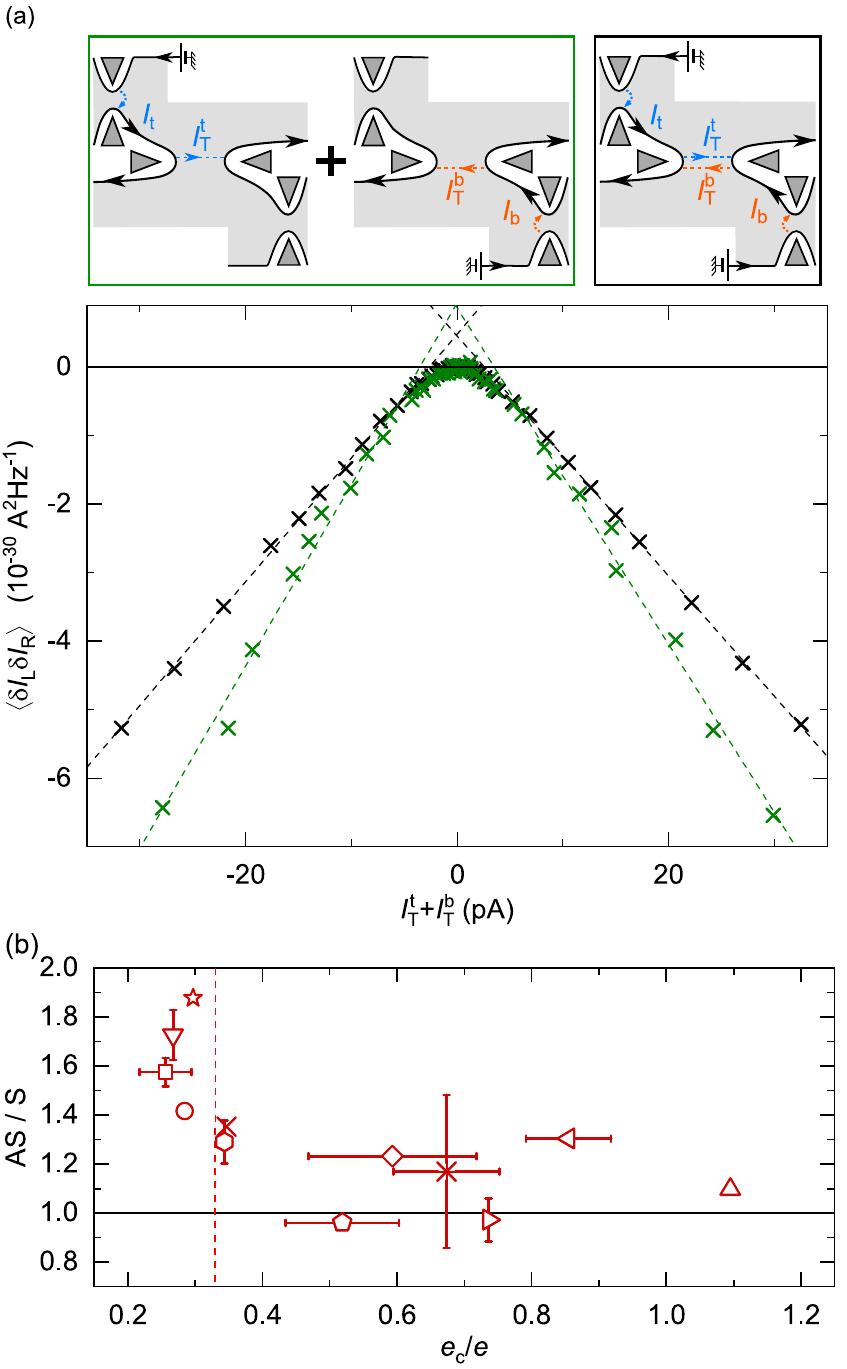}
	\caption{Discriminating anyons braiding from Andreev mechanisms.
	(a) Cross-correlations with QPC$_\mathrm{c}$ set to $e_\mathrm{c}\approx e_\mathrm{t,b}\approx e/3$.
	The data (symbols, corresponding to those in Fig.~\ref{Fig1s3robustness}) are displayed as a function of the sum $I_\mathrm{T}^\mathrm{t}+I_\mathrm{T}^\mathrm{b}$ of tunneling currents from the top and bottom sources (see the inset; arrows indicate the sign $+$ convention).
	Black crosses are obtained with two symmetric incident beams $I_\mathrm{t}\approx I_\mathrm{b}$. 
	Green crosses show the sum of independent measurements $\langle \delta I_\mathrm{L} \delta I_\mathrm{R} \rangle(I_\mathrm{t})+\langle \delta I_\mathrm{L} \delta I_\mathrm{R} \rangle(I_\mathrm{b})$, using either QPC$_\mathrm{t}$ or QPC$_\mathrm{b}$ as a single source, vs $I_\mathrm{T}^\mathrm{t}(I_\mathrm{t})+I_\mathrm{T}^\mathrm{b}(I_\mathrm{b})$.
	Dashed lines are linear fits.
	The significantly larger slope in the asymmetric case (green) rules out Andreev processes, and is consistent with the predicted anyon exchange mechanism. 
	(b) Symbols represent the ratio (AS/S) of the slopes $\langle \delta I_\mathrm{L} \delta I_\mathrm{R} \rangle / (I_\mathrm{T}^t+I_\mathrm{T}^b)$ between asymmetric (one source at a time, AS) and symmetric (two sources, S) incident quasiparticle beams versus separately characterized $e_\mathrm{c}/e$.
	Error bars encompass the difference between values extracted at negative and positive applied voltages for the same device setting.
	A ratio close to unity is found for $e_\mathrm{c}\gtrsim0.5>e_\mathrm{t,b}\approx e/3$, where the charge mismatch favors Andreev processes.}
	\label{Fig1s3asym}
\end{figure}

\subsection{Distinguishing anyon braiding and Andreev mechanisms} 
A distinctive feature of the unconventional anyon braiding mechanism, contrasting with the Andreev process, is that different incident quasiparticles do not contribute independently to the cross-correlations.
A straightforward test confirming this discriminating property is displayed in Fig.~\ref{Fig1s3asym}.

As graphically illustrated in Fig.~\ref{Fig1s3asym}(a), we compare, on the one hand, the sum of the cross-correlation signals measured alternatively with a single active source ($V_\mathrm{t}\neq0$ with $V_\mathrm{b}=0$, and $V_\mathrm{b}\neq0$ with $V_\mathrm{t}=0$; green) with, on the other hand, the cross-correlations measured when both sources are symmetrically biased ($V_\mathrm{t}=V_\mathrm{b}\neq0$; black). 
For Andreev processes the two match, as previously observed \cite{Glidic2022Andreev1s3}.
This is not the case when the underlying mechanism is the unconventional anyon exchange phase.
A representative comparison is displayed in Fig.~\ref{Fig1s3asym}(a) for the analyzer set in the WBS regime ($\tau_\mathrm{c}\simeq0.83$ with $e_\mathrm{c}\simeq0.34e$) where the anyon exchange mechanism is expected.
The marked difference between symmetric (black) and fully asymmetric (green) incident quasiparticle beams confirms that the underlying mechanism is not the Andreev process.

Figure~\ref{Fig1s3asym}(b) presents a systematic comparison for different analyzer tunings along the crossover between unconventional anyon exchange and Andreev mechanisms.
It is quantified by the displayed asymmetric to symmetric ratio `AS/S' between fitted linear slopes (e.g.\ dashed lines in Fig.~\ref{Fig1s3asym}(a)), plotted as a function of the parameter $e_\mathrm{c}/e$ driving the crossover.
When $e_\mathrm{c}\approx e/3$ (vertical dashed line), we systematically observe substantially larger cross-correlations in the asymmetric configuration, whereas for larger values of $e_\mathrm{c}$ the asymmetric to symmetric configurations ratio approaches $1$.
This signals a change of underlying transport mechanism when increasing $e_\mathrm{c}$, providing experimental support to the theoretical expectations of a crossover from unconventional exchange to Andreev processes.

The important dependence in the symmetry between incident beams of dilute quasiparticles, specifically observed when $e_\mathrm{c}\approx e/3$, constitutes a second qualitative marker of the unconventional exchange phase of the quasiparticle.

\begin{figure}[ht]
	\centering
	\includegraphics[width=8.6cm]
	{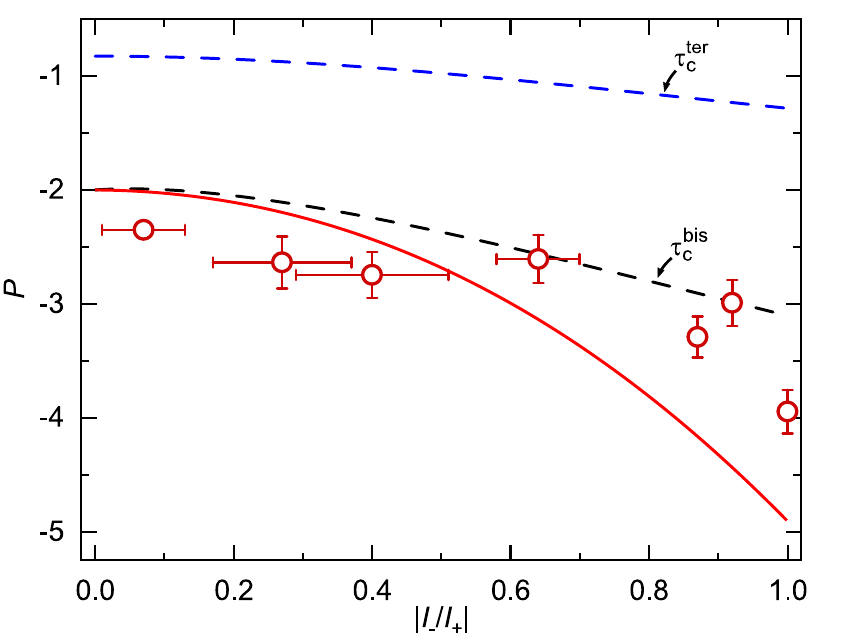}
	\caption{$P$ vs sources imbalance at $\nu=1/3$.
	Symbols display the effective Fano factor $P$ as a function of the relative difference in incident currents $I_-/I_+$, for the same device tuning. 
	Horizontal error bars encompass variations in $I_-/I_+$ over the range of applied voltages. 
	Vertical error bars represent the difference between values of $P$ separately extracted at negative and positive voltages. 
	The prediction of Eq.~\ref{EqPBraid} is shown as a red continuous line.
	The alternative predictions shown as dashed lines involve the different definitions $\tau_\mathrm{c}^\mathrm{bis}$ (black) and $\tau_\mathrm{c}^\mathrm{ter}$ (blue) (see Table~\ref{tab:FanoValues} and Appendix F).}
	\label{Fig1s3PvsAsym}
\end{figure}

\subsection{$P$ versus $I_-/I_+$}

We now confront quantitatively $P(I_-/I_+)$ data and the Eq.~\ref{EqP} prediction.

The experimental values of $P$ obtained for the analyzer QPC$_\mathrm{c}$ set to $\tau_\mathrm{c}\simeq0.7$ with $e_\mathrm{c}\simeq0.3e$ are displayed vs $I_-/I_+$ in Fig.~\ref{Fig1s3PvsAsym}.
For each data point, the ratio $V_\mathrm{t}/V_\mathrm{b}$ is kept fixed while sweeping $V_\mathrm{t,b}$.
Note that the variation of $|I_-/I_+|$ during each sweep, represented by the horizontal error bars, results from the unequal evolutions of $\tau_\mathrm{t}(V_\mathrm{t})$ and $\tau_\mathrm{b}(V_\mathrm{b})$.
The theoretical prediction of Eq.~\ref{EqP} is shown as a red continuous line.
For a more complete assessment, the predictions for $P_\mathrm{thy}^\mathrm{WBS}$ with the alternative definitions $\tau_\mathrm{c}^\mathrm{bis}$ and $\tau_\mathrm{c}^\mathrm{ter}$ (see Table~\ref{tab:FanoValues}) are also displayed as, respectively, black and blue dashed lines.

Theory predicts weak changes of $P$ at low $|I_-/I_+|$, progressively becoming stronger for higher asymmetries, consistent with experimental observations.
The expected ratio $P_\mathrm{thy}^\mathrm{WBS}(1)/P_\mathrm{thy}^\mathrm{WBS}(0)\simeq2.5$ is in order-of-magnitude agreement with the experimental value $P(1)/P(0)\sim1.5$.

Overall, the observed reasonable agreement between data and theory further corroborates the underlying presence of anyons of fractional exchange phase.

\section{Anyon signatures at $\nu=2/5$}

\begin{figure}[ht]
	\centering
	\includegraphics[width=8.6cm]
	{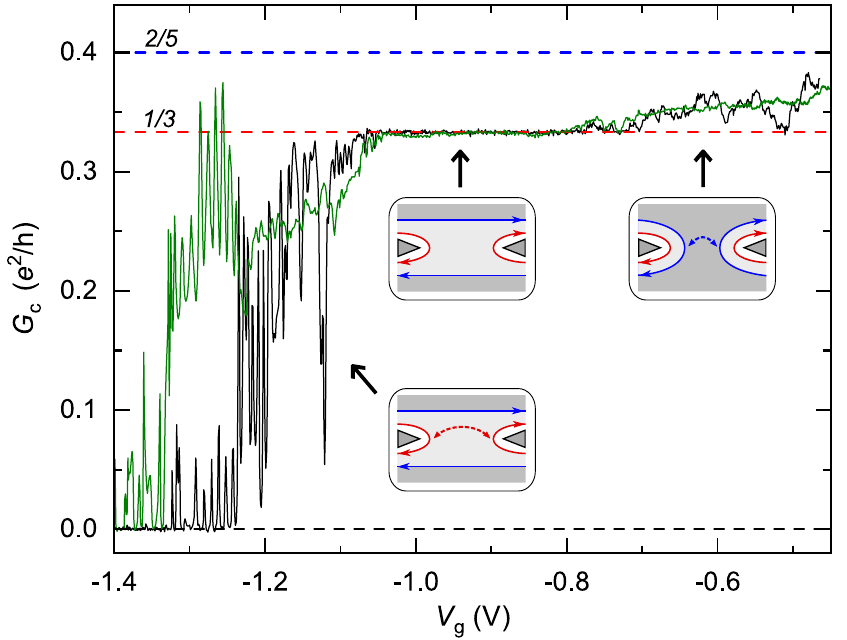}
	\caption{Differential conductance $G_\mathrm{c}$ through the analyzer QPC$_\mathrm{c}$ at $\nu=2/5$ as a function of the applied gate voltage $V_\mathrm{g}$ (detailed features may vary with overall device configuration).
	The black and green continuous lines display measurements, respectively, in the absence of a dc bias ($V_\mathrm{t}=V'_\mathrm{t}=V_\mathrm{b}=0$) and in the presence of a direct dc voltage bias ($V_\mathrm{t}=V'_\mathrm{t}=-90\,\mu$V, $V_\mathrm{b}=0$).
	The robust $e^2/3h$ plateau, where the absence of excess noise was separately checked, ascertains the sequential channel opening illustrated schematically.
	Note the relatively weak voltage bias dependence at $G_\mathrm{c}>e^2/3h$, when the inner (outer) edge channel is partially (fully) transmitted.
	}
	\label{Fig2s5GvsVg}
\end{figure}

\subsection{Edge structure}
At $\nu=2/5$, two adjacent channels are predicted to propagate in the same direction along each edge.
Quasiparticles of charge $e^*=e/5$ have been observed along the inner channel of conductance $\nu_\mathrm{eff}e^2/h$ with $\nu_\mathrm{eff}=1/15$ \cite{Reznikov_es5_1999}.
These quasiparticles are predicted to have a fractional exchange phase $\theta=3\pi/5$ ($\theta=2\pi\Delta$ with $\Delta=(e^*/e)^2/2\nu_\mathrm{eff}$, see e.g.\ \cite{Schiller2022scaldimFano}).

This edge structure is first attested by the dependence $G_\mathrm{c}(V_\mathrm{g})$ of the differential conductance $G_\mathrm{c}$ across the analyzer QPC$_\mathrm{c}$ with the voltage $V_\mathrm{g}$ applied to the metallic split gates controlling this constriction.
Figure~\ref{Fig2s5GvsVg} shows $G_\mathrm{c}(V_\mathrm{g})$ measured both at zero dc bias voltage (black line) and at $90\,\mu$V (green line).
The robust intermediate plateau at $G_\mathrm{c}=e^2/3h$ corresponds to the full transmission of the outer edge channel, of conductance $e^2/3h$, and the total reflection of the inner edge channel. 
For less negative $V_\mathrm{g}$, the higher $G_\mathrm{c}>e^2/3h$ reflects the subsequent opening of the inner edge channel of present interest.
The sequential, separated opening of the two channels is confirmed by the absence of excess noise when applying a dc voltage bias to QPC$_\mathrm{c}$ set on the $e^2/3h$ plateau.
The current transmission ratio of the inner edge channel at $G_\mathrm{c}\geq e^2/3h$ hence reads $(G_\mathrm{c}-e^2/3h)/(e^2/15h)$ (corresponding to $\tau_\mathrm{c}$ given by Eq.~\ref{EqTauC} in that direct voltage bias case).
Note that $G_\mathrm{c}$ does not reach the maximum value of $(2/5)e^2/h$ (horizontal blue dashed line) as it is not possible to fully open the inner channel across any of the QPCs. 
The maximum inner channel transmission achieved is $\tau_\mathrm{c}\approx0.9$ (see the inset in Fig.~\ref{Fig2s5tunneling}(d)). 
We refer to Appendix E for checks of the chirality of the electrical current in the central part of the device. 

In the presence of two channels co-propagating along each edge, a transfer (tunneling) of charges between adjacent channels along the source-analyzer paths could occur. 
As this results in an additional negative contribution to the measured cross-correlations $\langle \delta I_\mathrm{L}\delta I_\mathrm{R}\rangle$, its amplitude is carefully calibrated (see Appendix D). 
The tunneling current, made of $e/3$ quasiparticles as predicted \cite{Wen_LongPaperTLLFQHE_1992}, approaches at most $20\%$ of the injected inner channel current. 
A systematic procedure is set to estimate and subtract the tunneling current contribution to the cross-correlation signal (see Appendix D).
Importantly, the qualitative markers of braiding are not affected by this contribution; only the quantitative value of $P$ is modified, by at most $20\%$.

\begin{figure*}[t!]
	\centering
	\includegraphics[width=17.6cm]
	{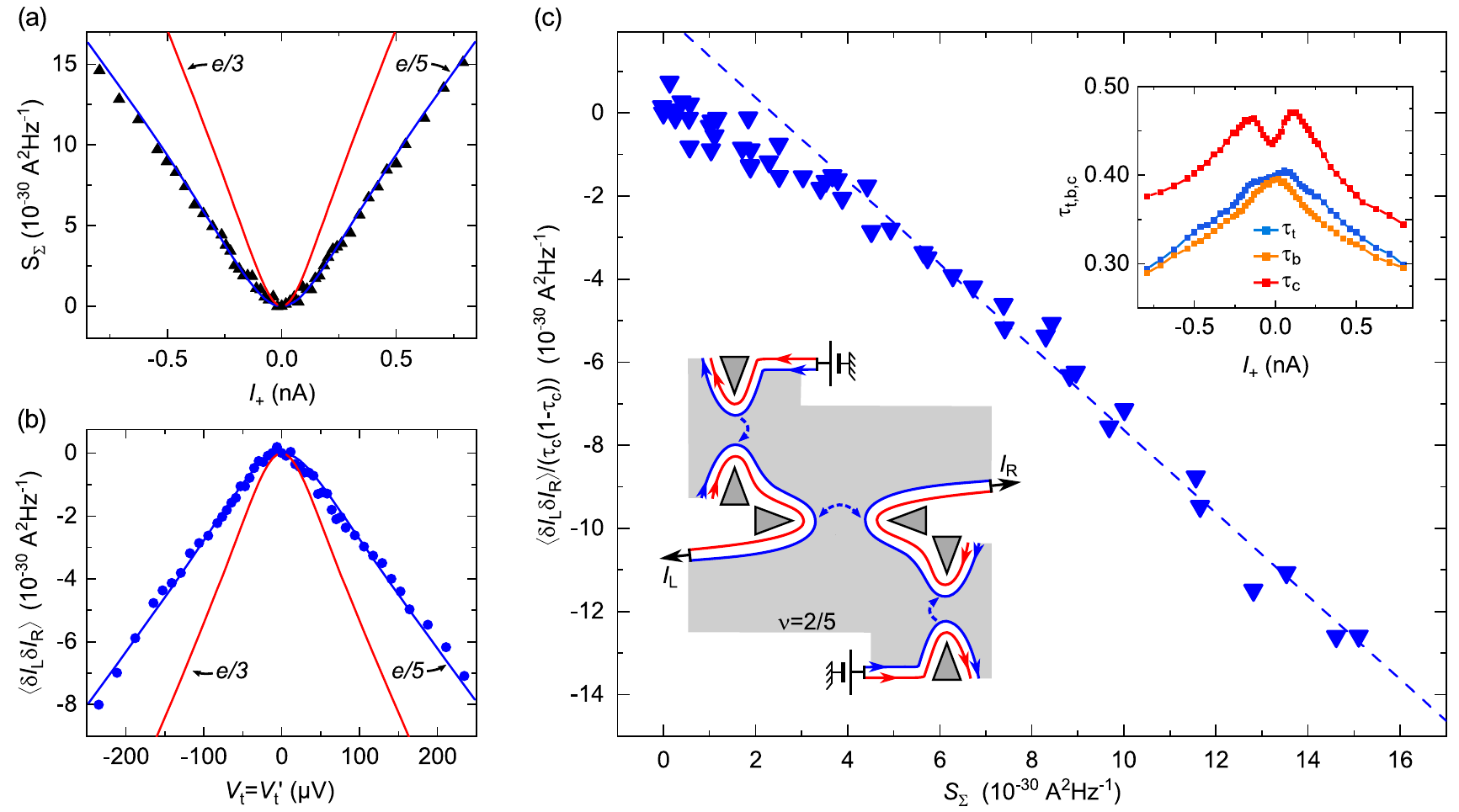}
	\caption{Cross-correlation signature of anyons in the inner edge channel at $\nu=2/5$. 
	The sources are symmetrically voltage biased ($V_\mathrm{t}=V_\mathrm{b}$, $V_\mathrm{t}'=0$) except for the separate analyzer characterization in (b) where $V_\mathrm{t}=V_\mathrm{t}'$ ($V_\mathrm{b}=0$) implements a direct voltage bias.
	(a),(b) Shot noise characterization of the charges $e_\mathrm{t,b}$ emitted from the sources (a), and of the tunneling charge $e_\mathrm{c}$ across the analyzer (b).
	The shot noise data (symbols) are compared with the predictions for $e^*=e/5$ (blue lines) and $e/3$ (red lines).
	(c) Experimental determination of $P$.
	The normalized cross-correlation data, from which the estimated inter-channel tunneling contribution (see Appendix~D) is removed, are plotted as symbols as a function of the same $S_\Sigma$ also shown in (a).
	$P\simeq-1.0$ is obtained from a linear fit of the slope (blue dashed line).
	Inset: simultaneous measurements of $\tau_\mathrm{t,b,c}$.
	}
	\label{Fig2s5repres}
\end{figure*}

\subsection{Representative anyon signature with symmetric $e/5$ sources}
Apart from inter-channel tunneling, the QPCs characterization and $P$ extraction procedures are similar to those at $\nu=1/3$, as illustrated in Fig.~\ref{Fig2s5repres}.
A charge $e_\mathrm{t,b}\simeq e/5$ for the quasiparticles emitted by the sources is obtained simultaneously with the measurement of $P$, by comparing the noise sum $S_\Sigma$ with the standard shot noise expression Eq.~\ref{EqSN} (see Fig.~\ref{Fig2s5repres}(a) for the simultaneous characterization of the sources).
Note that $S_\Sigma$ is not directly impacted by inter-channel tunnelings, as these processes preserve the overall current downstream the sources.
The characterization of $e_\mathrm{c}\simeq e/5$ is performed separately, analogous to $\nu=1/3$, from the cross-correlations measured in the presence of a direct voltage bias applied to QPC$_\mathrm{c}$ (see Fig.~\ref{Fig2s5repres}(b)).
Note that, unexpectedly, the noise sum $S_\Sigma$ (not shown in (b), see Fig.~\ref{fig:AutoVsCross}(c) in Appendix) is far from negligible although no voltage bias is applied to the sources in this configuration.
This might be related to a non-local heating, with most likely little impact on our conclusions (see Appendix E for further discussion).
The extraction of $P$ from the slope of the cross-correlation signal normalized by $\tau_\mathrm{c}(1-\tau_\mathrm{c})$ versus $S_\Sigma$ is shown in the main panel in Fig.~\ref{Fig2s5repres}(c).
We obtain in this representative example $P\simeq-1.0$ (dashed line, we extract $P\simeq-1.07$ from the raw data including inter-channel tunneling).
Qualitatively, the observed negative $P$ at symmetry indicates an unconventional anyon exchange phase.
Quantitatively, this is a markedly weaker value than the observed $P\approx-2$ at $\nu=1/3$.

\begin{figure}[htb]
	\centering
	\includegraphics[width=8.6cm]
	{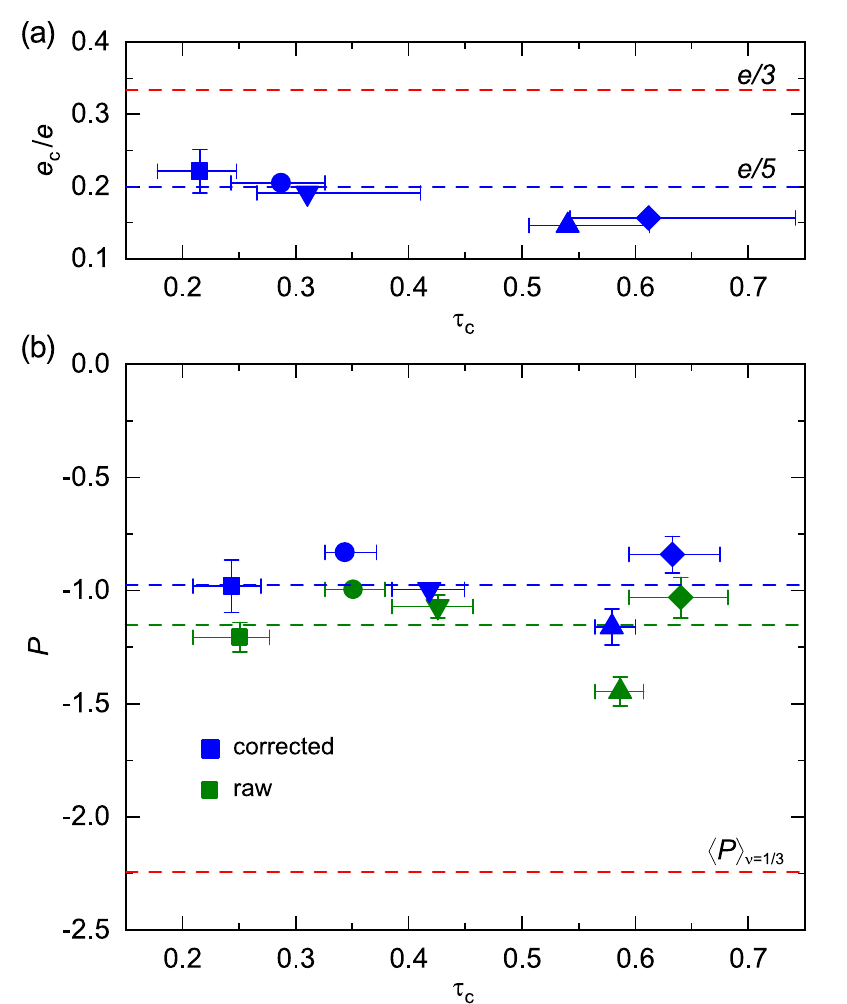}
	\caption{Effective Fano factor for $e/5$ quasiparticles at $\nu=2/5$.
	(a) Separately characterized analyzer tunneling charge $e_\mathrm{c}$ versus $\tau_\mathrm{c}$.
	(b) Experimental value of $P$ with (blue symbols) and without (green symbols) correcting for inter-channel tunneling.
	The same symbol as in (a) represents an identical device tuning ($\tau_\mathrm{c}$ changes due to a different biasing).
	The horizontal dashed lines indicate the corresponding mean values $\langle P \rangle \simeq -0.97$ (blue) and $\langle P \rangle \simeq -1.15$ (green), together with the $\nu=1/3$ observation $\langle P \rangle \simeq -2.24$  (red). 
	Error bars are displayed in (a) and (b) when larger than the symbols.
	Horizontal error bars display the variation of $\tau_\mathrm{c}$ during the measurement.
	Vertical error bars show the difference between values obtained separately for negative and positive applied voltages.}
	\label{Fig2s5Psym}
\end{figure}

\subsection{$P$ for symmetric $e/5$ quasiparticles sources}
Here we recapitulate the experimental effective Fano factor $P$ obtained for $e/5$ quasiparticles on five different device configurations (see Fig.~\ref{Fig2s5Psym}).

The analyzer displays a rather stable characteristic charge of $e_\mathrm{c}\approx e/5$, adapted to investigate the statistics of the corresponding quasiparticles, over a relatively broad explored range $\tau_\mathrm{c}\in[0.2,0.8]$ (see Fig.~\ref{Fig2s5Psym}(a)).
Note that for each tuning of QPC$_\mathrm{c}$, the sources QPC$_\mathrm{t,b}$ require gate voltage adjustments in order to preserve their symmetry.
In practice, the sources exhibit similar shot noise signatures of $e/5$ emitted quasiparticles, in comparably good agreement with Eq.~\ref{EqSN} than in the representative Fig.~\ref{Fig2s5repres}(a).
Note that the transmissions across QPC$_\mathrm{t,b}$ remain here within the range $\tau_\mathrm{t,b}\in[0.25,0.5]$, away from the dilute quasiparticle source limit that is experimentally not accessible. 
The extracted values of $P$ are recapitulated in Fig.~\ref{Fig2s5Psym}(b), with symbols matching those in (a) for identical device configurations (the different $\tau_\mathrm{c}$ result from the different biasing of QPC$_\mathrm{c}$).
The blue symbols represent $P$ obtained from the corrected cross-correlation signal, from which the contribution of inter-channel tunneling was subtracted.
The green symbols are the values of $P$ extracted from the raw cross-correlations.
The effect of inter-channel tunneling remains relatively small ($\sim20\%$) with respect to the overall value of $P$, and it does not introduce any noticeable trend.
Similarly to $\nu=1/3$, $P$ does not exhibit a significant dependence on $\tau_\mathrm{c}$.
However, in contrast, no crossover to a different Andreev-like mechanism could develop as there is here no mismatch between $e_\mathrm{t,b}$ and $e_\mathrm{c}$.
The simple observation of negative cross-correlations (much higher than from inter-channel tunnelings) thus points to a unconventional anyon exchange phase for the investigated $e/5$ quasiparticles.
Note that an exploration of the influence of an asymmetry between sources, used at $\nu=1/3$ to distinguish with Andreev physics, is here impeded by the high minimum values of experimentally accessible $1-\tau_\mathrm{t,b}\gtrsim0.5$, for which an applied asymmetry corresponds to a complex combination of incident quasiparticles and direct voltage bias. 

Quantitatively, we find an average value of $\langle P\rangle\simeq -0.97$ represented by a blue horizontal dashed line in Fig.~\ref{Fig2s5Psym}(b) ($\langle P\rangle\simeq -1.15$ from the raw data including inter-channel tunneling).
The theory developed for Laughlin fractions \cite{rosenow_collider_2016}, and recently extended to a non-Abelian channel \cite{lee2022nonabelian}, does not yet fully encompass hierarchical states such as $\nu=2/5$.
Nevertheless, assuming that the outer channel of conductance $e^2/3h$ can be ignored,
the same prediction $P=-4\Delta/(1-4\Delta)$ applies with the corresponding scaling dimension of the $e/5$ quasiparticles $\Delta=(e^*/e)^2/2\nu_\mathrm{eff}=0.3$ \cite{lee2022anyonshotnoise}.
The resulting $P=6$ is however much larger than observed and, intriguingly, positive.
The culprit for the sign change in this generalized prediction is not the cross-correlations, which remain negative as measured, but a differential transmission $\tau_\mathrm{c}$ becoming negative for dilute beams of such quasiparticles \cite{rosenow_collider_2016}.
Here we observe conventional, positive transmissions (see inset in Fig.~\ref{Fig2s5repres}(c)).
The important role of $\tau_\mathrm{c}$ in the theoretical value of $P$, with even more drastic consequences than at $\nu=1/3$, impedes the extraction of quantitative information on the specific anyon exchange phase.
It remains that the observation of strong negative cross-correlations constitutes a qualitative marker of unconventional exchange statistics.

\section{Conclusion}

Noise evidence of exotic anyon braiding statistics for fractional quasiparticles of charge $e/3$ and $e/5$ are observed in a source-analyzer setup.
This signature holds provided the analyzer QPC favors the transmission of the same type of quasiparticles as those emitted at the sources, and for relatively weak inter-channel tunnelings along the source-analyzer paths.
Different values for the cross-correlation effective Fano factor $P\approx-2$ and $P\approx-1$ are obtained, respectively, for $e/3$ quasiparticles at $\nu=1/3$ and $e/5$ quasiparticles along the inner channel at $\nu=2/5$ (in contrast with $P\approx0$ observed at $\nu=2$, see Appendix C).
It is tempting to attribute this difference to the distinct predicted exchange phases $\pi/3$ and $3\pi/5$.
However, the quantitative connection to $P$ is not direct, but involves the dependence of the analyzer transmission $\tau_\mathrm{c}$ on the voltages $V_\mathrm{t,b}$ used to generate the quasiparticles.
In practice, as generally observed experimentally in the fractional quantum Hall regime \cite{Chang2003,Heiblum_EdgeProbeTopo_2020}, the transmission across QPCs does not follow the predicted voltage bias dependence, which impedes any quantitative information on the exchange phase beyond its unconventional character.
A promising alternative to overcome this limitation is to combine such a source-analyzer setup with a quantum circuit implementation of Luttinger liquids \cite{safi_mappingTLL_2004,morel2021fractionalization} where QPCs are found to accurately follow the theoretical predictions \cite{Jezouin2013,Anthore2018}.

\begin{acknowledgments} 
This work was supported by the European Research Council (ERC-2020-SyG-951451), the European Union's Horizon 2020 research and innovation program (Marie Sk\l{}odowska-Curie Grant Agreement No.~945298-ParisRegionFP), the French National Research Agency (ANR-16-CE30-0010-01 and ANR-18-CE47-0014-01) and the French RENATECH network.

We thank A.~Kolli and C.~Loup-Forest for their help in the $\nu=2$ measurements, and C.~Mora, H.-S.~Sim, D.C.~Glattli, C.~Han, T.~Jonckheere, D.~Kovrizhin, and I.~Safi for helpful discussions. 

O.M.\ and P.G.\ performed the experiment and analyzed the data with inputs from A.Aa., A.An., C.P. and F.P.;
A.C., and U.G. grew the 2DEG;
A.Aa, F.P., O.M.\ and P.G.\ fabricated the sample;
Y.J.\ fabricated the HEMT used in the cryogenic noise amplifiers;
F.P., O.M.\ and P.G.\ wrote the manuscript with contributions from A.Aa., A.An., C.P.\ and U.G.;
A.An.\ and F.P.\ led the project.\\

\noindent\textit{Note added.}-- Our results are consistent with the independent investigation by M.~Ruelle \textit{et al.} \cite{ruelle_coll1s3-2s5_2022} submitted simultaneously with the present work. 
Relatively small quantitative discrepancies in $P$ for the inner channel at $\nu=2/5$ may be attributed to a different normalization procedure, where the measured $S_\Sigma$ here, see Eq.~\ref{EqP}, is replaced by the ideal Poissonian noise $2(e/5)I_+$ in \cite{ruelle_coll1s3-2s5_2022}.
After writing this manuscript, we also became aware of the related work \cite{lee2022anyonshotnoise}, supporting anyon braiding statistics at $\nu=1/3$ from a quantitative analysis of the auto-correlations in a single source - analyzer setup.

\end{acknowledgments}

\section*{Appendix A: Device} 
The measurements are performed on the same device previously used to evidence Andreev-like processes \cite{Glidic2022Andreev1s3}.
Note that the Ga(Al)As heterojunction hosting the 2D electron gas is similar to that used in the pioneer `collider' experiment \cite{bartolomei_anyons_2020}, and is grown in the same MBE chamber at a different time. 

The nanofabrication followed standard e-beam lithography steps (see Methods in \cite{Glidic2022Andreev1s3} for further details and large scale pictures of the sample): 
1) Ti-Au mark deposition through a PMMA mask.
2) Wet mesa etching in a solution of H$_3$PO$_4$/H$_2$O$_2$/H$_2$O through a positive ma-N 2403 mask.
3) Contact Ohmic deposition of Ni-Au-Ge through a PMMA mask, followed by a 440°C annealing for 50s.
4) Al gate deposition through a PMMA mask.
5) Deposition of Ti-Au bonding ports and large scale interconnects through a PMMA mask.

\begin{figure}[ht]
\centering
	\includegraphics[width=8.6cm]{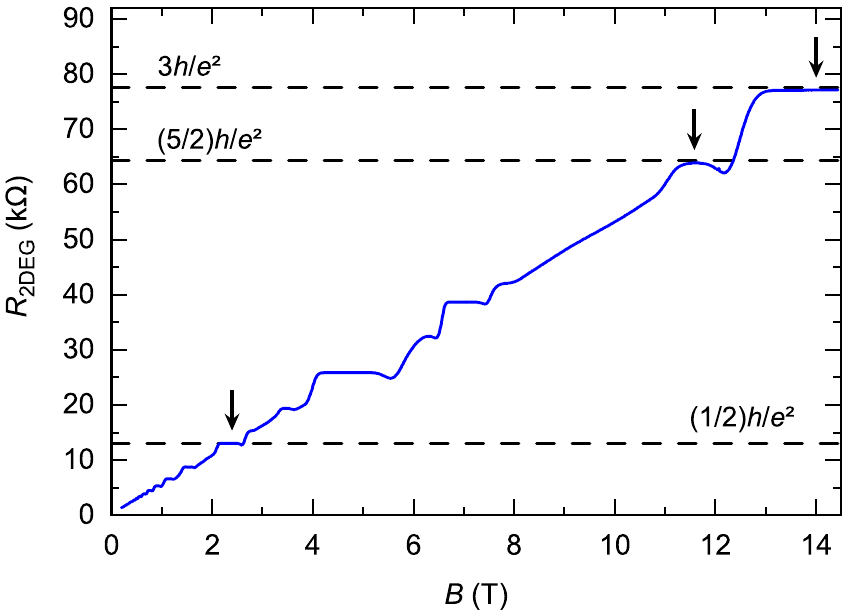}
	\caption{
	Quantum Hall resistance plateaus.
	Two-wire resistance between a Ohmic contact and cold grounds measured as a function of the applied perpendicular magnetic field $B$ at low temperature ($T\sim100$\,mK). Dashed lines show the fractional quantum Hall resistances $h/\nu e^2$ for the investigated fractions $\nu=1/3,$ $2/5$ and 2.
	Vertical arrows indicate the magnetic field at which the measurements are performed.
	}
	\label{fig:RvsB}
\end{figure}

Figure~\ref{fig:RvsB} shows the 2-wire measurement at $T\sim100$~mK of the resistance between an Ohmic contact and cold grounds as a function of the magnetic field (a fixed wiring and filtering resistance of 10.35\,k$\Omega$ is subtracted).
The experiments are performed in the center of the plateaus at $\nu=1/3$, $2/5$ and 2, at the values indicated by vertical arrows.
Note that the effective magnetic field range for the plateaus is, on the one hand, slightly reduced by density gradients over the sample and, on the other hand, increased by the temperature reduction when probing the anyon statistics.

The Ohmic contacts have a large perimeter of approximately $200\,\mu$m with the 2D electron gas to ascertain an essentially perfect electrical connection (usually already achieved for perimeters of about $10\,\mu$m with our recipe).
The ohmic contact quality together with the robustness of edge transport chirality in the 2DEG are attested by \textit{(i)} the accurate resistance of the quantum Hall plateaus, \textit{(ii)} the absence of current reflected from ohmic contacts connected to a cold ground and \textit{(iii)} the absence of excess shot noise when closing all the QPCs and applying a voltage bias.

\section*{Appendix B: Experimental setup}

\subsection*{Measurement setup}
Measurements are performed in a cryofree dilution refrigerator, where the sample is connected through electrical lines including several thermalization and filtering stages (see \cite{Iftikhar2016} for comprehensive technical details). 

Auto- and cross-correlations of current fluctuations are measured near 1\,MHz using two home-made cryogenic HEMT amplifiers (see supplemental material in \cite{Jezouin2013b} for further information) respectively connected to the $L$ and $R$ ports of the sample as schematically depicted in Fig.~\ref{FigSample}(c). 

All other measurements are performed with standard lock-in techniques, using ac modulation of rms amplitude below $k_\mathrm{B}T/e$ and at frequencies lower than 25\,Hz. 
The transmitted dc currents are obtained by integrating the corresponding lock-in differential signal with respect to the applied bias voltage (explicit expressions are provided in Methods in \cite{Glidic2022Andreev1s3}).

\subsection*{Thermometry}

At $T>40$\,mK the electronic temperature is determined using a calibrated RuO$_2$ thermometer, thermally anchored to the mixing chamber of the dilution refrigerator. 
In this range, the thermal noise of the sample changes linearly with $T$ confirming both the RuO$_2$ thermometer calibration and the good thermalization of the charge carriers in the device.

To obtain an in-situ electronic temperature below 40\,mK, we measure the thermal noise and extrapolate the noise-temperature slope determined at higher $T$. 

\subsection*{Noise amplification chain calibration}

The gain factors $G^\mathrm{eff}_\mathrm{L,R,LR}$ relating the raw auto/cross-correlations with the power spectral density of current fluctuations of interest is calibrated in two steps.

First, the nearly identical tank circuits connected to the Ohmic contacts labeled $L$ and $R$ are characterized.
This is achieved by measuring the variation of the noise bandwidth of each of the tank circuits in parallel with the quantum Hall resistance of the sample at several filling factors, which informs on the parallel tank resistance $R_\mathrm{tk}\approx150\,\mathrm{k}\Omega$ and capacitance $C_\mathrm{tk}\approx135\,$pF.
The resonant frequency (0.86\,MHz) then provides the parallel tank inductance $L_\mathrm{tk}\approx250\,\mu$F.

Second, with our choice of noise integration bandwidth ($[0.84,0.88]$\,MHz at $\nu=2/5$ and $1/3$), we measure the slopes $s_\mathrm{tk}$ of raw integrated noise vs temperature at $T>40$\,mK (see Thermometry). 
The robust fluctuation-dissipation relation then gives the gain factors $G^\mathrm{eff}_\mathrm{L,R}=s_\mathrm{tk}/(4k_\mathrm{B}(1/R_\mathrm{tk}+\nu e^2/h))$, whereas the nearly identical tanks imply for the cross-correlation $G^\mathrm{eff}_\mathrm{LR}=\sqrt{G^\mathrm{eff}_\mathrm{L}G^\mathrm{eff}_\mathrm{R}}$. 
See \cite{Glidic2022Andreev1s3} for a more thorough presentation including checks with alternative methods.

\begin{figure*}[htb]
	\centering
	\includegraphics[width=17.6cm]
	{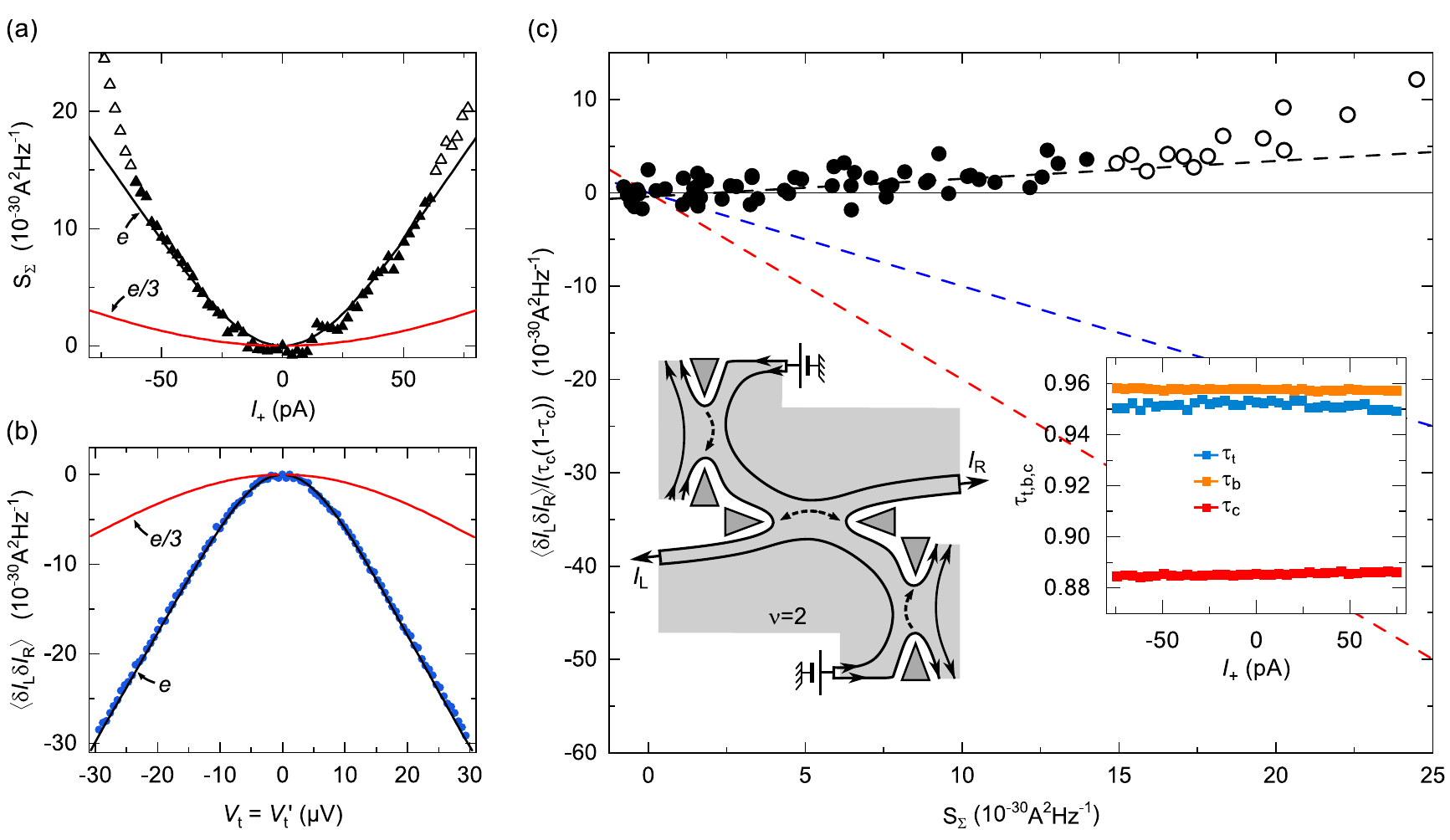}
	\caption{Cross-correlations at $\nu=2$ with symmetric sources. 
	All QPCs are set in the WBS regime ($\tau_\mathrm{t}\simeq\tau_\mathrm{b}\simeq 0.96$ and $\tau_\mathrm{c}\simeq 0.88$, see the inset in (c)) with $V_\mathrm{t}=V_\mathrm{b}$ and $V_\mathrm{t}'=0$, except for the analyzer characterization in (b) where $V_\mathrm{t}=V_\mathrm{t}'$ and $V_\mathrm{b}=0$ corresponding to a direct voltage bias applied to QPC$_\mathrm{c}$.
	(a),(b) Shot noise characterization of the tunneling charges $e_\mathrm{t,b}$ (a) and $e_\mathrm{c}$ (b).
	Symbols display the noise data, in close match with the prediction for a tunneling charge $e$ (black lines) at $S_\Sigma\lesssim15\,10^{-30}\,\mathrm{A}^2/$Hz (full symbols).
	(c) Cross-correlations measured in the symmetric sources-analyzer configuration (symbols) are plotted vs source shot noise $S_\Sigma$.
	A linear fit at $S_\Sigma<15\,10^{-30}\,\mathrm{A}^2/$Hz (black dashed line) gives $P\simeq+0.2$. 
	The red and blue dashed line represent the Fano factor obtained at $\nu=1/3$ and $\nu=2/5$, respectively. 
	Inset: transmission probabilities of top (blue symbols), bottom (orange symbols) and central (red symbols) QPCs as a function of $I_+$.
	}
	\label{fig:nu=2}
\end{figure*}

\section*{Appendix C: Cross-correlation investigation of the statistics at $\nu=2$}
As a counterpoint to anyons, we present here measurements of $P$ with the device set in the integer quantum Hall regime at filling factor $\nu=2$ ($B=2.4$\,T).
In this regime, electrons with a Fermi statistics are emitted at the source QPCs and transmitted across the analyzer QPC.
Note that interactions between the two co-propagating channels are predicted to drive a transition of the propagating excitations from Fermi quasiparticles to bosonic density waves, resulting in the emergence of positive cross-correlations in the sources-analyzer setup \cite{idrisov2022poscrossnu2}.
However, for the present short propagation distance of $1.5\,\mu$m and in the accessible small voltage bias range before artifacts develop $|V|\lesssim30\,\mu$V, the interactions between the two co-propagating channels are essentially negligible \cite{leSueur_relax_2010,Roulleau_LphiIQHE_2008}.
Note also that inter-channel tunneling is here completely negligible.

Figure~\ref{fig:nu=2} displays representative data obtained at $\nu=2$, with symmetric sources emitting in the outer edge channel toward the analyzer.
The procedure is identical to in the FQHE regime. 
The panels (a) and (b) show the tunneling charge characterization of the sources and analyzer, respectively, found to match the shot noise predictions for $e$ in both cases.
Note that a huge noise develops at high bias (emerging for the highest $I_+$ in (a)), thus limiting the investigated range.
The panel (c) represents the cross-correlation signal in the source-analyzer configuration with symmetric incident dilute beams, normalized by $\tau_\mathrm{c}(1-\tau_\mathrm{c})$ and plotted versus $S_\Sigma$. 
A linear fit of the data displayed as full symbols for which $S_\Sigma<15\,10^{-30}\,\mathrm{A}^2/$Hz (black dashed line) gives $P\simeq 0.2$.

Although not exactly null, $P$ is here very small with respect to the observed $P\simeq-2$ (red dashed line) and $P\simeq-1$ (blue dashed line) at $\nu=1/3$ and $2/5$, respectively.
The slight positive value may result from the essentially but not fully negligible inter-channel interactions, which progressively change the nature of electronic excitations along the sources-analyzer paths.
The present small and positive $\nu=2$ data hence corroborate the predicted link between negative cross-correlations and unconventional anyon statistics.

\begin{figure*}[htb]
	\centering
	\includegraphics[width=17.2cm]
	{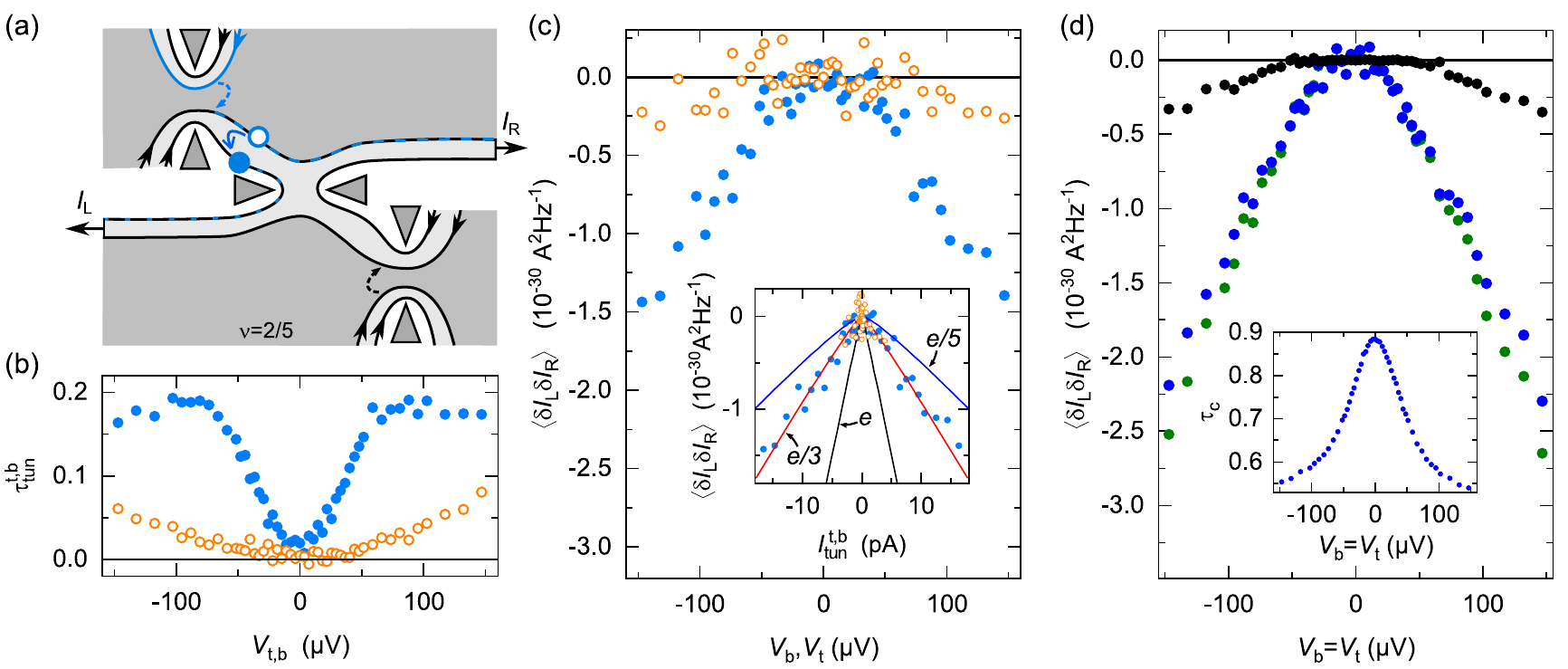}
	\caption{Inter-channel tunneling at $\nu=2/5$.
	(a) Schematics of inter-channel tunneling along the top source-analyzer path.
	For a separate characterization, QPC$_\mathrm{c}$ is set to $\tau_\mathrm{c}=0$ (and a full transmission of the outer channel).
	(b) Inter-channel tunneling fraction.
	Symbols display the ratio between tunneling and emitted current along the top ($\tau_\mathrm{tun}^\mathrm{t}$, blue) and bottom ($\tau_\mathrm{tun}^\mathrm{b}$, orange) paths, obtained at $\tau_\mathrm{c}=0$ from Eq.~\ref{EqTauICtun}.
	The displayed values approaching 20\% are the highest observed in all investigated configurations.
	(c) Inter-channel tunneling cross-correlations at $\tau_\mathrm{c}=0$.
	Symbols represent the separately obtained signals for tunnelings along the top (blue) and bottom (orange) paths, as a function of the dc bias voltage of the corresponding source $V_\mathrm{t}$ (with $V_\mathrm{b}=0$) and $V_\mathrm{b}$ (with $V_\mathrm{t}=0$), respectively.
	Inset: The same cross-correlations are plotted as a function of the corresponding inter-channel tunneling current $I_\mathrm{tun}^\mathrm{t,b}$ and compared with the shot noise predictions of Eq.~\ref{EqSN} (lines).
	(d) Cross-correlations measured with QPC$_\mathrm{c}$ tuned back to inner channel analyzer ($\tau_\mathrm{c}>0$), with both sources symmetrically biased ($V_\mathrm{t}=V_\mathrm{b}$) are shown as green symbols.
	Black symbols display the inter-channel tunneling contribution estimated from (c) (sum of data in (c) times $(1-\tau_\mathrm{c})^2$, see text).
	The resulting `corrected' cross-correlations (raw data reduced by tunneling estimate) are shown as blue symbols.
	Inset: simultaneously measured $\tau_\mathrm{c}$.
	}
	\label{Fig2s5tunneling}
\end{figure*}

\section*{Appendix D: Inter-channel tunneling at $\nu=2/5$}

As illustrated in Fig.~\ref{Fig2s5tunneling}(a), inter-channel tunnelings, if any, result in an additional negative contribution to the measured cross-correlations $\langle \delta I_\mathrm{L}\delta I_\mathrm{R}\rangle$, thereby impacting $P$.
Indeed, a tunnel-induced current fluctuation $\delta I$ in the outer channel is correlated to an opposite fluctuation $-\delta I$ in the inner channel.
With a downstream QPC$_\mathrm{c}$ of inner channel transmission ratio $\tau_\mathrm{c}$ and perfectly transmitted outer channel, the resulting total current fluctuation (summed over both channels) in the transmitted and reflected $L,R$ paths is $\delta I -\delta I\tau_\mathrm{c}$ and $-\delta I(1-\tau_\mathrm{c})$, respectively, corresponding to a cross-correlation signal of $-\delta I^2(1-\tau_\mathrm{c})^2$.
Note that the fluctuation $\delta I$ depends on the charge of the tunneling quasiparticles, here between two markedly different channels.
With the procedure described below, we find \textit{(i)} that inter-channel tunnelings can occur when some power is locally injected into the inner channel at the corresponding upstream source QPC (set at $0<\tau_\mathrm{t,b}<1$, but note that no tunneling is here observed at $\tau_\mathrm{t,b}=0$); \textit{(ii)} that the noise in the inter-channel tunneling current is consistent with the predicted tunneling charge of $e/3$ (determined by the local filling factor of $1/3$ of the incompressible stripe separating the two channels, see \cite{Wen_LongPaperTLLFQHE_1992}); but \textit{(iii)} that this contribution here remains relatively small with respect to the cross-correlation signal of present interest, generated at the analyzer QPC$_\mathrm{c}$.

Figure~\ref{Fig2s5tunneling} illustrates the experimental procedure to address inter-channel tunneling, in the device configuration where it is found to be the strongest, at $T\simeq25\,$mK.
The central QPC$_\mathrm{c}$ is first detuned from the inner channel analyzer operating point ($\tau_\mathrm{c}>0$) and set to the $e^2/3h$ plateau ($\tau_\mathrm{c}=0$).
In order to minimize any cross-talk artifacts, we change only the voltage applied to the QPC$_\mathrm{c}$ gate located the furthest away from the separately considered path (the gate along $I_\mathrm{L(R)}$ for the path originating from QPC$_\mathrm{b(t)}$, see Fig.~\ref{FigSample}(c)).
The differential tunneling transmission ratio $\tau_\mathrm{tun}^\mathrm{t(b)}$ of the inner channel current into the outer channel along the top (bottom) source-analyzer path simply reads, at $\tau_\mathrm{c}=0$ and for a sequential channel opening of the QPCs:
\begin{equation}
    \tau_\mathrm{tun}^\mathrm{t(b)}=\frac{\partial I_\mathrm{L(R)}/\partial V_\mathrm{t(b)}}{\partial (I_\mathrm{L}+I_\mathrm{R})/\partial V_\mathrm{t(b)}}.
    \label{EqTauICtun}
\end{equation}
As shown in Fig.~\ref{Fig2s5tunneling}(b), the tunneling ratio along the top path can here approach $20\%$ of the injected inner channel current (the maximum value observed in all the device configurations investigated), markedly higher than along the bottom path.
The simultaneously measured cross-correlations $\langle \delta I_\mathrm{L}\delta I_\mathrm{R}\rangle$ are displayed in Fig.~\ref{Fig2s5tunneling}(c) as a function of the applied voltage $V_\mathrm{t}$ (blue symbols) or $V_\mathrm{b}$ (orange symbols) in the main panel, and as a function of the dc inter-channel tunneling current $I_\mathrm{tun}^\mathrm{t}$ (blue) or $I_\mathrm{tun}^\mathrm{b}$ (orange) in the inset.
As seen in the inset, the cross-correlations resulting from inter-channel tunneling match the shot noise prediction of Eq.~\ref{EqSN} for the expected $e^*=e/3$.

QPC$_\mathrm{c}$ is then set back to the inner channel analyzer operating point $\tau_\mathrm{c}>0$ ($\approx0.9$ at zero bias here, see the inset in Fig.~\ref{Fig2s5tunneling}(d)) and the cross-correlation signal is measured in the presence of symmetric beams of quasiparticles generated at the source QPCs now simultaneously biased at the same $V_\mathrm{t}=V_\mathrm{b}$.
The green symbols in the main panel represent the raw signal, which includes the additional negative contribution from inter-channel tunnelings.
The estimate of this unwanted contribution (black symbols) is obtained by simply applying the reduction factor $(1-\tau_\mathrm{c})^2$ to the inter-channel cross-correlations previously measured at $\tau_\mathrm{c}=0$.
Note that inter-channel tunneling also changes the relation providing $\tau_\mathrm{c}$.
The impact is generally found to be relatively small (of at most $0.04$ at high bias for the present example), however Eq.~\ref{EqTauC} should be modified by substituting $(1-\tau_\mathrm{t(b)})$ with $(1-\tau_\mathrm{t(b)})(1-\tau_\mathrm{tun}^\mathrm{t(b)})$ to account for the reduction of the incident inner channel current.
In this work, we will extract the effective Fano factor $P$ from both the measured cross-correlation signal ignoring inter-channel tunnelings (green symbols), and by removing the estimated inter-channel tunneling contribution from the measurements (blue symbols).
Confronting the two obtained values of $P$ allows one to straightforwardly appreciate the relatively small influence of inter-channel tunnelings (see Fig.~\ref{Fig2s5Psym}(b)).

\section*{Appendix E: Supplementary data}

\begin{figure}[ht]
	\centering
	\includegraphics[width=8.6cm]
	{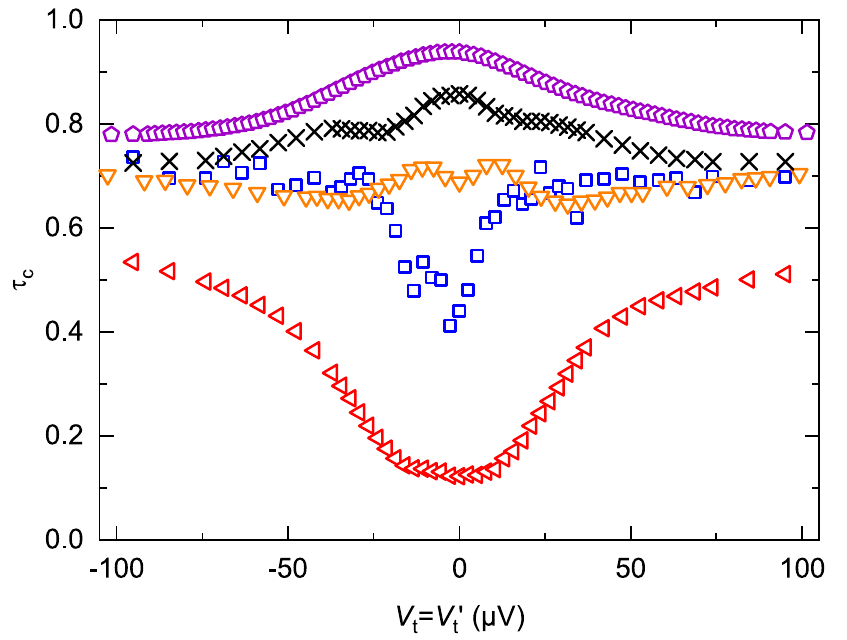}
	\caption{Analyzer QPC$_\mathrm{c}$ transmission $\tau_\mathrm{c}$ versus direct voltage bias $V_\mathrm{t}=V_\mathrm{t'}$ at $\nu=1/3$.
	Different symbols correspond to different tunings of QPC$_\mathrm{c}$.
	An identical device tuning to Fig.~\ref{Fig1s3robustness} is represented here by the same symbol.
	}
	\label{fig:tauCdependance}
\end{figure}

\subsection*{Bias dependence of QPC transmission at $\nu=1/3$}

In the FQHE regime, the current transmission ratio $\tau$ across a QPC is predicted to depend on bias voltage \cite{Kane&Fisher1992PRL,Fendley_fraccharge_1995}.
This energy dependence on the analyzer transmission $\tau_\mathrm{c}$ influences the quantitative prediction for $P$, as discussed in the main text.
However, experimentally, the QPC transmissions are generally found in disagreement with the expected voltage biased dependence (see e.g.\ \cite{Chang2003}).
The transmission vs direct voltage bias characterization of the analyzer QPC$_\mathrm{c}$ at $\nu=1/3$ is shown for a broad range of tuning in Fig.~\ref{fig:tauCdependance}.
In the WBS regime of present main interest, we find that the transmission $\tau_\mathrm{c}$ is reduced, getting further away from the ballistic limit as the direct voltage bias is increased.
Similar observations were made by other teams (see e.g.\ \cite{Heiblum_EdgeProbeTopo_2020}), as well as for both source QPCs and for the outer $e^2/3h$ channel of QPC$_\mathrm{c}$ at $\nu=2/5$.
This contrasts with the prediction of a transmission approaching unity as the bias is increased \cite{Kane&Fisher1992PRL,Fendley_fraccharge_1995}.

\begin{figure}[htb]
	\centering
	\includegraphics[width=8.6cm]
	{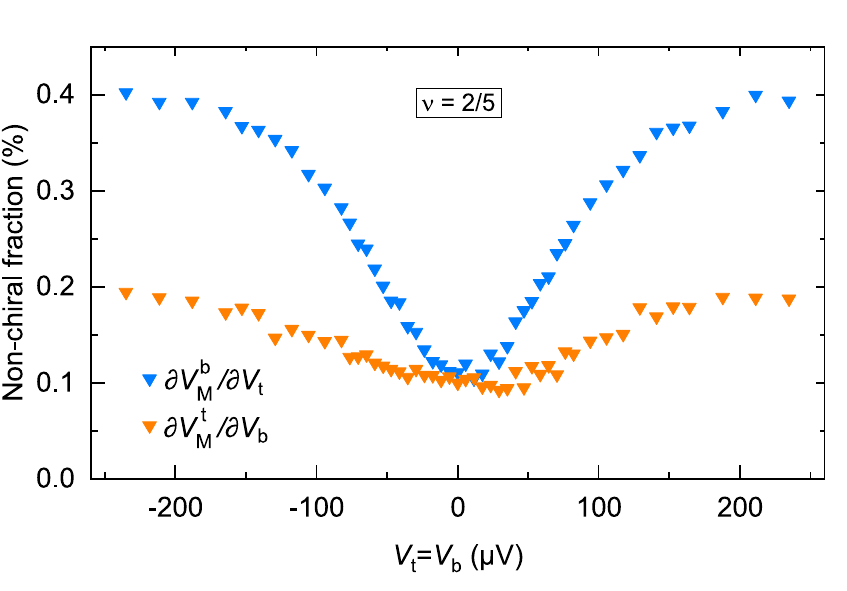}
	\caption{
	Non-chiral signal at $\nu=2/5$.
	The displayed differential emitted/detected current (voltage) ratios $\partial V_\mathrm{M}^\mathrm{t(b)}/\partial V_\mathrm{b(t)}$ should be null for a perfectly chiral system. 
	}
	\label{fig:Chirality}
\end{figure}

\subsection*{Transport chirality}

The quantum Hall chirality of the electrical current is systematically obeyed at the level of the large Ohmic contacts.
Nevertheless, we find that small but discernible deviations can develop at the heart of the device for the less robust $\nu=2/5$ fractional quantum Hall state.

The local chirality is controlled by checking that the signals $\partial V_\mathrm{M}^\mathrm{t}/\partial V_\mathrm{b}$ and $\partial V_\mathrm{M}^\mathrm{b}/\partial V_\mathrm{t}$ are null, as $V_\mathrm{M}^\mathrm{t(b)}$ should be disconnected from $V_\mathrm{b(t)}$ by chirality whatever the device tuning.
This is always the case at experimental accuracy at $\nu=2$ and $\nu=1/3$, but a small unexpected signal is found at $\nu=2/5$ as illustrated in Fig.~\ref{fig:Chirality}.
If we consider that the non-chiral signal originates solely from the inner channel current, the relevant non-chiral fraction is enhanced by a factor of $6$ (i.e.\ $(2/5)/(1/15)$), up to $2.5\%$ in the present representative example and at the most 3\% in the worst case investigated.

\begin{figure*}[htb]
	\centering
	\includegraphics[width=17.6cm]
	{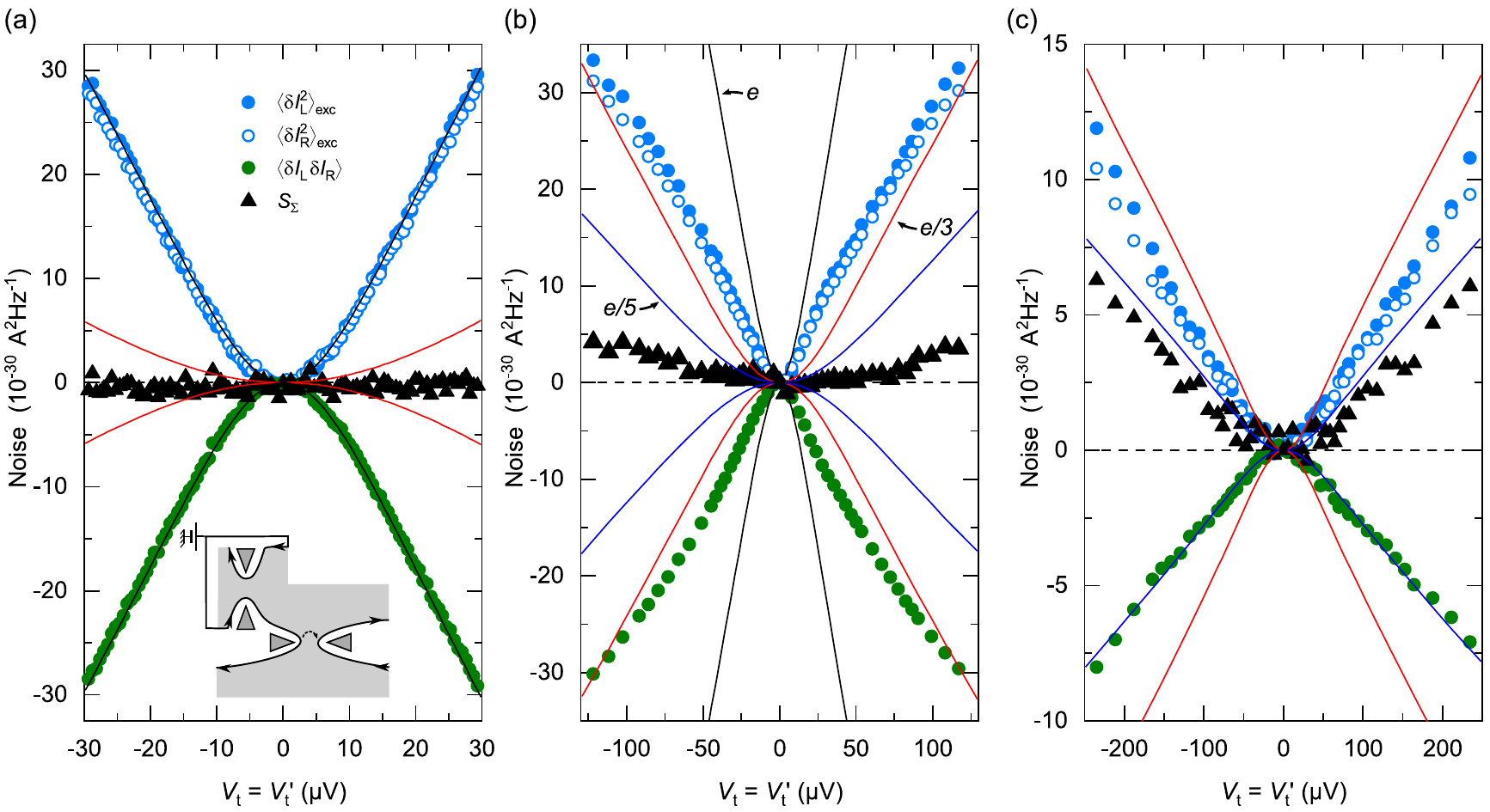}
	\caption{Auto/Cross-correlations comparison, performed in QPC$_\mathrm{c}$ characterization at filling factor $2$ (a), $1/3$ (b) and $2/5$ (c) as a function of the direct applied voltage $V_\mathrm{t}=V_\mathrm{t}'$ ($V_\mathrm{b}=0$). 
	Full and open blue discs display the auto-correlation signal from port $L$ and $R$ respectively.
	Green discs represents the simultaneously measured cross-correlations.
	$S_\Sigma$ is plotted as black triangles.
	Continuous lines display ($\pm1\times$) the predictions of Eq.~\ref{EqSN} for a charge $e^*=e$ (black), $e/3$ (red) and $e/5$ (blue).}
	\label{fig:AutoVsCross}
\end{figure*}

\subsection*{Auto- vs cross-correlations in QPC$_\mathrm{c}$ characterization}
It was pointed out that the cross-correlations could provide a more robust probe than the auto-correlations for the shot noise characterization of the tunneling charge across QPCs in the FQHE regime \cite{Kapfer2018}. 
Here, we compare auto- and cross-correlation signals measured with a direct voltage bias applied to QPC$_\mathrm{c}$, during the separate $e_\mathrm{c}$ characterization.

Figure~\ref{fig:AutoVsCross} shows measurements of the auto-correlations and cross-correlations, as well as the corresponding noise sum $S_\Sigma$, obtained at $\nu=2$ (a), $1/3$ (b) and $2/5$ (c).
The cross-correlations (green circles) correspond to the previously displayed representative data in Figs.~\ref{fig:nu=2}(b), \ref{Fig1s3repres}(b) and \ref{Fig2s5repres}(b), now completed with coincidental measurements of $\langle \delta I_\mathrm{L}^2\rangle$ (full blue circles) and $\langle \delta I_\mathrm{R}^2\rangle$ (open blue circles), and with $S_\Sigma$ defined in Eq.~\ref{EqSsigma}.
Continuous lines represent the shot noise predictions of Eq.~\ref{EqSN} (with a $-1$ factor when negative) at the measured $T\simeq35$\,mK for $e^*=e$ (black), $e/3$ (red) and $e/5$ (blue).
At $\nu=2$ a canonical behavior is observed, with opposite auto- and cross-correlations both corresponding to $e^*=e$ and resulting vanishing noise sum $S_\Sigma\simeq0$.
At $\nu=1/3$, small but discernible deviations from $S_\Sigma=0$ develop at high voltage bias, which signal the emergence of small differences between auto- and cross-correlations.
These are attributed to a non-local heating of the source QPCs resulting in a noise increase such as the delta-$T$ noise \cite{Lumbroso2018,Sivre2019,Larocque_dTnoise_2020} (see Methods in \cite{Glidic2022Andreev1s3} for a specific investigation on the same sample).
At $\nu=2/5$, the sum noise $S_\Sigma$ is far from negligible, which might be related to the non-local heating observed at $\nu=1/3$ although stronger. 

One may wonder if the unexpected $S_\Sigma$ signal observed at $\nu=2/5$ with a direct voltage biased applied to QPC$_\mathrm{c}$ could impact our conclusions.
As argued below, we believe it is unlikely.
First, the doubts that this discrepancy casts on $e_\mathrm{c}$ would not directly impact the extracted value of $P$ (see Eq.~\ref{EqP}).
Second, we point out that the cross-correlations chosen to characterize $e_\mathrm{c}$ were previously established to be more reliable than the auto-correlations \cite{Kapfer2018}.
This is even more true in the present source-analyzer setup where the incident current noise can be enhanced by a non-local heating of the sources. 
Finally, if such unexpected increase of $S_\Sigma$ were to occur also in the main sources-analyzer configuration, the absolute value of $P$ involving $S_\Sigma$ in the denominator would be reduced but would not vanish (Eq.~\ref{EqP}).   
Moreover, in the sources-analyzer configuration a voltage bias is applied to the sources (as opposed to $e_\mathrm{c}$ characterization), which is expected to suppress the effect of a local heating on $S_\Sigma$ (see Eq.~\ref{EqSN}). 
Accordingly, the reliability of $S_\Sigma$ with voltage biased sources is supported by the similar $e_\mathrm{t,b}$ extracted when detuning the analyzer QPC$_\mathrm{c}$ to $\tau_\mathrm{c}=0$ and biasing the source QPCs one at a time (data not shown).
These considerations suggest that the unexpectedly high $S_\Sigma$ observed when applying a direct voltage bias to QPC$_\mathrm{c}$ is likely to have a moderate impact on $e_\mathrm{t,b}$ and $P$, without qualitative consequences on the present anyon statistics investigation.

\section*{Appendix F: $P_\mathrm{thy}^\mathrm{WBS}$ with alternative $\tau_\mathrm{c}$}

In this section, we provide the analytical expressions for the theoretical predictions of $P_\mathrm{thy}^\mathrm{WBS}$ as defined by Eq.~\ref{EqP}, but using $\tau_\mathrm{c}^\mathrm{bis}$ and $\tau_\mathrm{c}^\mathrm{ter}$ instead of $\tau_\mathrm{c}$.
These predictions, valid for all QPCs in the WBS limit and for large source voltages with respect to $k_\mathrm{B}T/e^*$, are shown in Table~\ref{tab:FanoValues} for $I_-=0$ and $I_-=I_+$.

First, we consider $\tau_\mathrm{c}^\mathrm{bis}(I_-,I_+)\equiv \tau_\mathrm{c}(I_-=0,I_+)$.
This corresponds to the choice of normalization made in \cite{rosenow_collider_2016}.
The effective Fano factor in the WBS regime $1-\tau_\mathrm{c}^\mathrm{bis}\ll1$ and at large bias voltage then reads:
\begin{equation}
\begin{split}
        P_\mathrm{thy,bis}^\mathrm{WBS}=& \frac{\langle \delta I_\mathrm{L}\delta I_\mathrm{R}\rangle}{2 e^* I_+ (1-\tau_\mathrm{c}^\mathrm{bis})}=\left[1+\left(\frac{|I_-|}{|I_+|\tan(2\pi\Delta)}\right)^2 \right]^{2\Delta-1}\\
        &\times \Big(-\frac{4\Delta}{1-4\Delta}\cos X\\
        &+\frac{|I_-|}{|I_+|}
        \left[ \tan(2\pi\Delta)+\frac{\tan^{-1}(2\pi\Delta)}{(1-4\Delta)}\right]
        \sin X \Big),
        \label{EqPbis}
\end{split}
\end{equation}
with $X\equiv (4 \Delta-2) \arctan\big(|I_-|/|I_+| \tan^{-1}(2\pi\Delta)\big)$.

This expression reduces Eq.~\ref{EqPBraid} at $I_-=0$, since in that limit $\tau_\mathrm{c}^\mathrm{bis}=\tau_\mathrm{c}$. 
For $\Delta=1/6$ at $\nu=1/3$, it gives $P_\mathrm{thy,bis}^\mathrm{WBS}(0)=-2$ and $P_\mathrm{thy,bis}^\mathrm{WBS}(I_-/I_+=1)\simeq-3.1$ as shown in Table~\ref{tab:FanoValues}.
Equation~\ref{EqPbis} for $\Delta=1/6$ corresponds to the black dashed line in Fig.~5 of the present article, and to the continuous line shown in the bottom right panel in Fig.~3 in \cite{rosenow_collider_2016}.

Second, we consider $\tau_\mathrm{c}^\mathrm{ter}\equiv \tau_\mathrm{t}^{-1}\partial V_\mathrm{L}/\partial V_\mathrm{t}'$, which is the transmission ratio for thermal excitations with the sources in the WBS limit.
This normalization choice was made in \cite{lee2022nonabelian} (see the alternative Fano factor called $P_\mathrm{ref}$).
The corresponding effective Fano factor in the WBS regime $1-\tau_\mathrm{c}^\mathrm{ter}\ll1$ and at large bias voltage reads \cite{lee2022nonabelian}:
\begin{equation}
\begin{split}
        P_\mathrm{thy,ter}^\mathrm{WBS}=  P_\mathrm{thy,bis}^\mathrm{WBS}(I_-/I_+)\frac{\sin(4\pi\Delta)}{4\pi\Delta}.
        \label{EqPter}
\end{split}
\end{equation}
At $\Delta=1/6$ ($\nu=1/3$), the reduction factor is $P_\mathrm{thy,ter}^\mathrm{WBS}/ P_\mathrm{thy,bis}^\mathrm{WBS}\simeq0.41$ and Eq.~\ref{EqPter} gives $P_\mathrm{thy,ter}^\mathrm{WBS}(0)\simeq-0.83$ and $P_\mathrm{thy,ter}^\mathrm{WBS}(I_-/I_+=1)\simeq-1.28$ as shown in Table~\ref{tab:FanoValues}.
Equation~\ref{EqPter} at  $\Delta=1/6$ corresponds to the blue dashed line in Fig.~5 of the present article, and to the black continuous line in Fig.~4 in \cite{lee2022nonabelian}.


\end{document}